\documentclass[]{interact}

\usepackage{epstopdf}
\usepackage[caption=false]{subfig}
\usepackage{caption}
\DeclareCaptionLabelSeparator{none}{ }
\usepackage[doublespacing]{setspace}
\usepackage[natbibapa,nodoi]{apacite}
\usepackage{multirow}
\usepackage{tabu}
\usepackage{xcolor}
\usepackage{csquotes}

\theoremstyle{plain}

\theoremstyle{definition}

\theoremstyle{remark}

\begin{document}

\articletype{ARTICLE TEMPLATE}

\title{Critical Point Calculations by Numerical Inversion of Functions}

\author{
\name{C. N. Parajara\textsuperscript{a}, G. M. Platt\textsuperscript{b}\thanks{G. M. Platt. Email: gmplatt@furg.br}, F. D. Moura Neto\textsuperscript{a}, M. Escobar\textsuperscript{b} and G. B. Libotte\textsuperscript{a}}
\affil{\textsuperscript{a}Department of Computational Modeling, Rio de Janeiro State University, Nova Friburgo, Brazil; \textsuperscript{b}School of Food and Chemistry, Federal University of Rio Grande, Santo Ant\^{o}nio da Patrulha, Brazil}
}

\maketitle

\begin{abstract}
In this work, we propose a new approach to the problem of critical point calculation, based on the formulation of 
\citet{HeidemannKhalil}. This leads to a $2 \times 2$ system of nonlinear algebraic equations in temperature and molar volume, which makes possible the prediction of critical points of the mixture through an adaptation of the technique of inversion of functions from the plane to the plane, proposed by \citet{bib:malta1993geometria}. The results are compared to those obtained by three methodologies: \textit{(i)} the classical method of \citet{HeidemannKhalil}, which uses a double-loop structure, also in terms of temperature and molar volume; \textit{(ii)} the algorithm of \citet{Dimitrakopoulos}, which employs a damped Newton algorithm and \textit{(iii)} the methodology proposed by \citet{nichita2010}, based on a stochastic algorithm. The proposed methodology proves to be robust and accurate in the prediction of critical points, as well as provides a global view of the nonlinear problem.
\end{abstract}

\begin{keywords}
critical point; numerical inversion of functions; functions from the plane to the plane
\end{keywords}

\section{Introduction}

Critical point calculations have been extensively studied in Chemical Engineering field\textcolor{black}{, since important thermodynamic phenomena and separation operations are dependent on the accurate values for the critical coordinates; for instance, the behavior of oil reservoirs \citep{bib:peters1998} and supercritical extraction processes \citep{raeissi2004phenomenon}}. \textcolor{black}{In this scenario, two methodologies have been largely employed in the last decades: the method of  \citet{hicks} and the algorithm of \citet{HeidemannKhalil}. The methodology proposed by Heidemann and Khalil is particularly interesting, since the calculation of critical points is represented by a nonlinear algebraic system, to be solved for critical temperature and molar volume in a double-loop structure. Thus, the large quantity of algorithms capable to solve nonlinear sytems can be used (and tested) to solve the problem, such as Newton-type algorithms \citep{bib:deuflhard2011newton} or even stochastic optimization methods (also called metaheuristics; in this case, the nonlinear algebraic system is usually converted into a scalar merit function, to be minimized) \citep{nichita2010}. Both methodologies are extremely useful in most cases (for the Newton-type methods, when the nonlinearity of the problem is not sufficient to demand extremely precise initial estimates). On the other hand, there are some occasions where more robust methods must be employed.}   

In many occasions,  thermodynamic systems exhibit more than one critical point. In this sense, several strategies to obtain---in a robust way---all critical points of such  systems have been developed. \textcolor{black}{In these occasions, classical Newton-type algorithms and metaheuristics may exhibit difficulties to obtain multiple critical points.} Considering this situation, \citet{stradi} proposed an interval Newton-generalized bisection algorithm to solve the critical point problem. Their algorithm exhibits mathematical and computational guarantees to obtain all critical points. \textcolor{black}{ More recently, \citet{sidky2016} employed a polynomial homotopy continuation algorithm--- also a robust method--- in the calculation of critical points. Previously, \citet{wang1999} used a homotopy-based method to obtain the critical loci of binary mixtures.}

In this work we propose the solution of this challenging problem with the use of a robust algorithm, based on the numerical inversion of functions from the plane to the plane, developed by \citet{bib:malta1996numerical}. This algorithm has already been applied in some Chemical Engineering problems, such as double azeotropy \citep{GuedesPlattMouraNeto} and double retrograde vaporization \citep{artigoGustavo,industrial2018}. \textcolor{black}{In addition, this algorithm is also capable to describe --- in a geometrical way --- the nonlinear behavior of the critical point calculation, allowing a better understanding of the complex nature of the problem and the related thermodynamic models.}

This work is organized as follows. In the next section, we formulate 
the problem. In Section~\ref{sec:num_methodology}, the numerical 
methodology used to obtain critical points is described. Then, in 
Section~\ref{sec:results} we present the results obtained for \textcolor{black}{four binary mixtures}, comparing them with results from the literature. 
In the last section some conclusions and final remarks are drawn.

\section{Problem Formulation} \label{sec:formulation}

\textcolor{black}{We will use the critical point calculation as formulated by \citet{HeidemannKhalil}, as detailed in the Appendix.} \textcolor{black}{Using this approach, the critical point problem can be expressed by  \citep{rochoczetal,StockflethDohrn}}:
\begin{subequations}
\label{duas}
\begin{equation}
\sum\limits_{j=1}^{c}\sum\limits_{i=1}^{c}\left(\dfrac{\partial^2 A}{\partial n_i \partial n_j}\right)_{T_0,V_0,n_{0,l\neq j,i}}\Delta n_i \Delta n_j = 0\;,
\label{a1}
\end{equation}
\vspace{-1em}
\begin{equation}
\sum\limits_{k=1}^{c}\sum\limits_{j=1}^{c}\sum\limits_{i=1}^{c}\left(\dfrac{\partial^ 3 A}{\partial n_i\partial n_j\partial n_k}\right)_{T_0,V_0,n_{0,l\neq k,j,i}}\Delta n_i\Delta n_j \Delta n_k  = 0\;.
\label{a2}
\end{equation}
\end{subequations}
\textcolor{black}{In the previous equations, $A$ refers to the Helmholtz free energy, $c$ is the number of components in the mixture, $T$ is the absolute temperature, $V$ is the molar volume and $n$ refers to the number of mols.}

According to \cite{HeidemannKhalil}, \cite{StockflethDohrn} and \cite{Dimitrakopoulos}, the derivatives of Helmholtz free energies $A$ can be substituted using standard thermodynamic relations, by expressions involving the derivatives of the fugacities $\hat{f}_i$ of the components in the mixture:
\begin{equation}\label{fugacidade_helmholtz}
\left( \dfrac{\partial ^2 A}{\partial n_i \partial n_j}\right)_{T,V,n_{l\neq i,j}} = RT\left(\frac{\partial\ln{\hat{f}_i}}{\partial n_j}\right)_{T,V,n_{,l\neq j}}\;.
\end{equation}



 Let $\bold{Q}$ be a $c\times c$ matrix of compositional derivatives, whose elements $q_{ij}$ are given by
\begin{equation}\label{qp}
q_{ij} = \left( \dfrac{\partial\ln{\hat{f}_i}}{\partial n_j}\right)_{T_0,V_0,n_{0,l\neq j}}\;.
\end{equation}
Also, let $\bold{\Delta}\bold{n}$ denote 
the vector of the variation on the number of mols, $\bold{\Delta}\bold{n}=(\Delta n_1,\Delta n_2,
\ldots, \Delta n_c )^t$. 
%
By Equations~(\ref{fugacidade_helmholtz}) and (\ref{qp}), 
 Equations~(\ref{duas}) can be rewritten as
\begin{subequations}
\label{dois}
\begin{equation}\label{eq1_sistema}
\bold{\Delta} \bold{n}^{t}\bold{Q}\bold{\Delta} \bold{n}=0 \;,
\end{equation}
\begin{equation}\label{eq2_sistema}
\sum\limits_{k=1}^{c}\sum\limits_{j=1}^{c}\sum\limits_{i=1}^{c}\left(\dfrac{\partial^
2 \ln{\hat{f}_i}}{\partial n_j\partial n_k}\right)_{T_0,V_0,n_{0,l\neq k,j}}\Delta n_i\Delta n_j\Delta n_k = 0\;.
\end{equation}
\end{subequations}

The classical critical conditions of \citet{HeidemannKhalil} require
furthermore 
that $\bold{Q}$ be positive semi-definite and, therefore, $\det \bold{Q}=0$.
In this case, the homogeneous system $\bold{Q}\bold{\Delta} \bold{n} = \bold{0}$ admits nontrivial solutions (in fact, an infinite number of solutions). One way out is to choose arbitrarily one element of the vector $\bold{\Delta} \bold{n}$  
 (typically, one chooses the first element equal to 1). The other components of the vector are then calculated.

Given the composition of a mixture, 
the nonlinear system to be solved for $(T,V,\bold{\Delta} \bold{n})$ can be stated as 
\begin{subequations}
\label{sistema_final}
\begin{eqnarray}
\det \mathbf{Q}&=&0\:, \label{9a}
\\
\sum\limits_{k=1}^{c}\sum\limits_{j=1}^{c}\sum\limits_{i=1}^{c}\left(\dfrac{\partial^
 2 \ln{\hat{f}_i}}{\partial n_j\partial n_k}\right)_{T,V,n_{l\neq k,j}}\Delta n_i\Delta n_j\Delta n_k & = & 0 \:,\label{9b}
 \\
 \bold{Q}\bold{\Delta} \bold{n}&=&\bold{0}\:. \label{9c}
\end{eqnarray}
\end{subequations}

Letting 
\begin{equation}
\bold{G}(T,V,\bold{\Delta} \bold{n}) = \left(
 \det \bold{Q}\:,\ \ 
 \sum\limits_{k=1}^{c}\sum\limits_{j=1}^{c}\sum\limits_{i=1}^{c}\left(\dfrac{\partial^
 2 \ln{\hat{f}_i}}{\partial n_j\partial n_k}\right)_{T,V,n_{l\neq k,j}}\Delta n_i\Delta n_j\Delta n_k\:, \ \ \bold{Q}\bold{\Delta} \bold{n}
\right)\:.
\end{equation}
The system can be restated as $\bold{G}(\bold{p})=\bold{q}$,
where $\bold{p}$ is a point in the domain,
 $\mathbf{G} = (G_1,\ldots, G_{c+2})$, $\bold{q}$ is the null vector,
 and $\bold{p} = (T,V,\bold{\Delta} \bold{n})$.

In the simulations, we consider mixtures with two components, $c=2$, 
therefore this setup
can be simplified a little bit.
In this case, Equation~(\ref{9c}) has
only one independent equation. Using the first one, and 
letting $\Delta n_1=1$, we get $\Delta n_2=-q_{11}/q_{12}$, {\em i.e.}
$\bold{\Delta} 
\bold{n}=(\Delta n_1,\Delta n_2)=(1, -q_{11}/q_{12})$.
Substituting these values in Equation~\ref{9b}, and defining function
$\bold{F}$ by the left hand side of Equations~\ref{9a} and \ref{9b},
\begin{eqnarray}
\bold{F}(T,V)=\left(\det \mathbf{Q}\:, 
\sum\limits_{k=1}^{c}\sum\limits_{j=1}^{c}\sum\limits_{i=1}^{c}\left(\dfrac{\partial^
 2 \ln{\hat{f}_i}}{\partial n_j\partial n_k}\right)_{T,V,n^0}\Delta n_i\Delta n_j\Delta n_k\right)
 \label{final2}
\end{eqnarray}
the equations of the critical thermodynamic points are written simply
as
\begin{eqnarray}
\bold{F}(T,V)=(0,0)\:, \label{final}
\end{eqnarray}
which is interpreted as computing the pre-image of a point, $(0,0)$,
by a nonlinear map from the plane to the plane, $\bold{F}$.

\section{Numerical Methodology} \label{sec:num_methodology}

As described in the previous section, the numerical modeling of 
critical points is represented by a $(c+2) \times (c+2)$  nonlinear 
algebraic system which can be converted to a $2 \times 2$ problem for binary mixtures. A huge quantity of numerical procedures can be 
applied in the solution of this kind of  system. Among these, we can 
mention Newton-type methods \citep{Dimitrakopoulos}, 
homotopy-continuation procedures \citep{wang1999,sidky2016} and 
interval methods \citep{stradi}. Here, we employ the methodology of 
numerical inversion of functions from the plane to the plane, proposed 
by \cite{bib:malta1993geometria}. The main purpose of this method is 
to numerically determine all pre-images of a given point $\bold{q}$ in 
the image of the function. It must be stressed that the problem 
formulation is the same employed by \citet{HeidemannKhalil}, but with 
a different numerical procedure to obtain the solution. 
\citet{HeidemannKhalil} uses a double-loop structure in terms of temperature and molar volume, solved by Newton or secant methods.

The original methodology of the inversion of functions from the plane to the plane is applicable to a specific group of functions, as pointed out by \cite{bib:malta1996numerical}. Here, we use some numerical techniques described by \cite{bib:malta1996numerical}, but not the whole methodology. First, we present some features of the global geometry of nonlinear functions by considering a simple example,
as a motivation to some calculations that are the base of
the methodology of the inversion of functions from the plane to the plane.
Next, the three major steps of the method will be described, which in general terms represent the obtaining of the critical set (in the domain) and the critical image (the image of the critical set in the image of $ \bold{F} $), the generation of the bank of solved points and the inversion of an arbitrary point.

\subsection{Global geometry of nonlinear functions: a one dimensional illustration}\label{31}

Here we illustrate, by means of a  simple one dimensional algebraic 
case, some features of 
how the global geometry of a nonlinear function $F$ 
impacts in the ensuing nonlinear equation $F(p)=q$, for the
determination of the solutions $p$, for  a given $q$. 

\begin{figure}
\centering
\includegraphics[width=\textwidth]{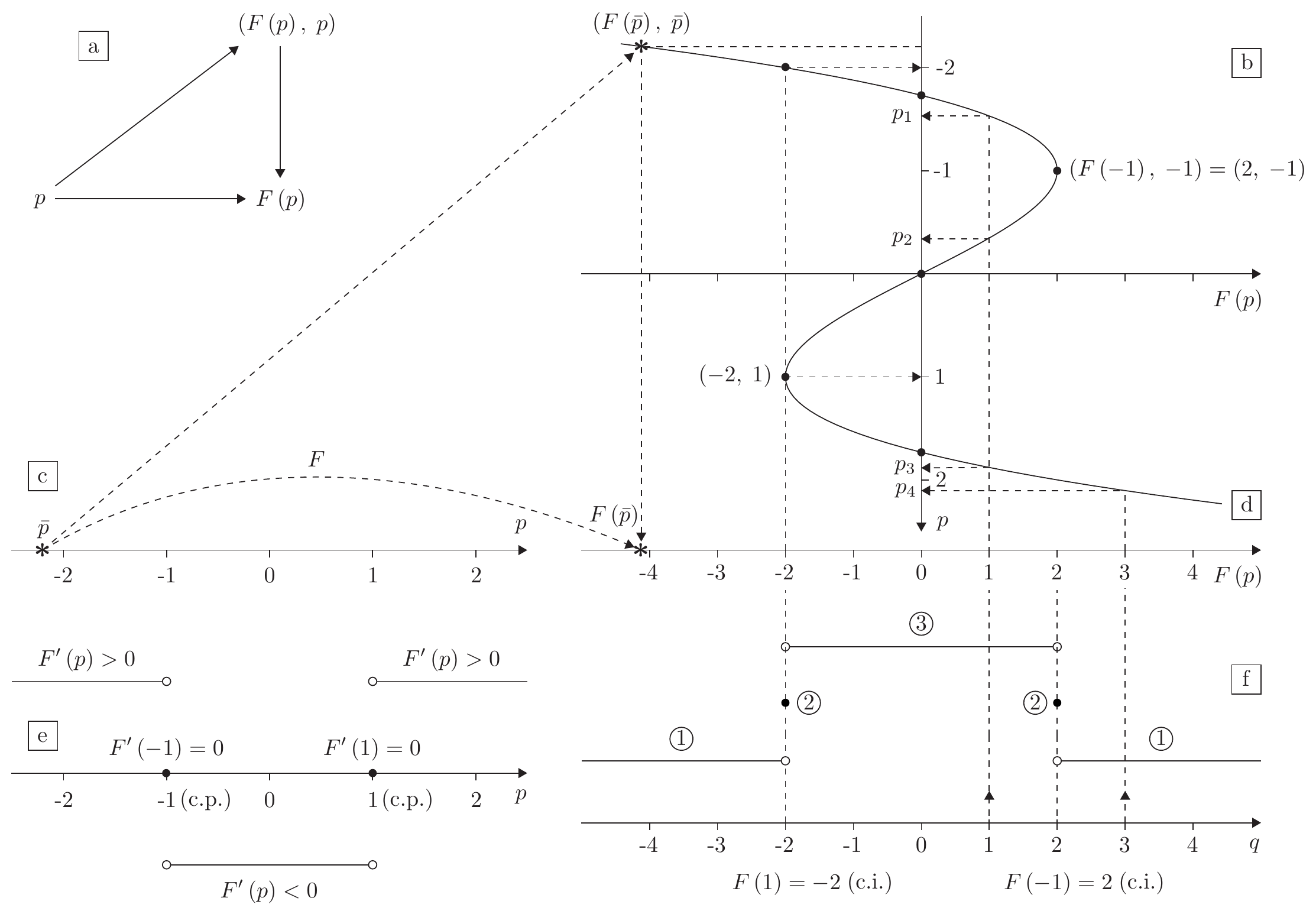}
\caption{ a) A commutative diagram showing the role of the domain, codomain and graph of a nonlinear function in the understanding of a 
nonlinear function. The domain 
is stretched, bended and folded to produce the graph. Next, the graph
is projected onto the codomain to complete the nonlinear mapping of the 
domain into the codomain (compare (a) with (c)-(b)-(d)); b) The graph of $F(p)=p^3-3p$; c) The domain of $F$; d) The codomain of $F$; e) The sign of $F'$ and the critical points of $F$; f) The number of solutions of the equation $F(p)=q$ depending on where $q$ lies, and the critical images.}
\label{fpoli}
\end{figure}
Let $F(p)=p^3-3p$. Figure~\ref{fpoli} shows several aspects of 
this function. The graph of $F$ is represented in Figure~\ref{fpoli}b, 
although the axes are rotated by $\pi/2$ in clockwise direction.
This way it is easier to understand what the nonlinear
 function makes out 
of the $p$-domain (Figure~\ref{fpoli}c)---the real line---when mapping
it to the $q$-codomain (Figure\ref{fpoli}d); beginning in the 
$p$-domain (Figure~\ref{fpoli}c),
it stretches, bends and folds it to construct the graph of $F$
(see Figure~\ref{fpoli}b) 
before embedding it into the $q$-codomain (Figure~\ref{fpoli}d), by an orthogonal projection.   
Figure~\ref{fpoli}a is just a sketch of all of this, showing a commutative diagram: either one takes the domain first to the graph and then projects onto the codomain, or one goes directly from the domain to the codomain. 

Now we illustrate the role that the 
critical points of a nonlinear function (in the mathematical sense, {\em i.e.}, points where $F'=0$)
have in determining the number of solutions for the equation. Their critical images are transition points where the number of solutions may change; otherwise the number of solutions is constant.
In the example, the critical points (c.p.) are the solutions of $F'(p)=3p^2-3=0$, {\em i.e.}, $p=\pm 1$, and the critical images (c.i.) are $F(-1)=2$ and $F(1)=-2$. Figure~\ref{fpoli}e shows the critical points, 
and the sign of $F'$, and Figure~\ref{fpoli}f shows the critical images and the number of solutions the nonlinear equation $F(p)=q$ has, depending on where $q$ lies. It is clear that the number of solutions change when $q$ crosses a critical image. One has:
\begin{itemize}
\item 3 solutions if $ -2<q<2$ (with $q= 1$, the solutions are 
$p_1\sim -1.53$, $p_2\sim-0.35$  and $p_3\sim 1.88$;)
\item
2 solutions if  $q=\pm 2$ (with $q= -2$, the solutions are $-2 $
and 1);
\item
one solution if $q<-2$  or $q>2$  (with $q= 3$, the solution is 
$p_4\sim 2.10 $).
\end{itemize}

\subsection{Generation of the critical set}

The first step of the methodology of the inversion of functions from the plane to the plane to solve $2\times 2$ systems 
of nonlinear equations is to determine the critical set of the function, in the mathematical sense, as explained next. A
point $\bold{p}$ is a critical point of  a function $\bold{F}$ if  $\det(\bold{J}(\bold{p})) = \bold{0}$, where $\bold{J}$ is the 
Jacobian of
 the function $\bold{F}$. The critical set of the function $\bold{F}$ 
 from the plane to the plane is a collection of critical curves (not  thermodynamic ones) formed by critical points of $\bold{F}$.

As pointed out by \cite{artigoGustavo}, in the original methodology proposed by \cite{bib:malta1996numerical}, a premise is that the functions must be \textit{nice}. Functions of this type have to
satisfy some specific requirements: (\textit{i}) $\mathbf{F}$ is proper; (\textit{ii}) the critical set of $\mathbf{F}$ is bounded; (\textit{iii}) the (mathematical) critical points are folds or cusps and; (\textit{iv}) the image of $\mathbf{F}$ is a normal set of curves. A detailed discussion of nice functions, folds and cusps,
as well as a very detailed characterization 
of the solutions of the equation $\bold{F}(\bold{p})=\bold{q}$,
for various values of $\bold{q}$,  is presented by \cite{bib:malta1993geometria}. Here, the assumption of a nice function is relaxed, but we must bear in mind that all the guarantees discussed in \cite{bib:malta1996numerical} are not valid in our computations.

\begin{figure}[!htbp]
\centering
\includegraphics[scale=0.9]{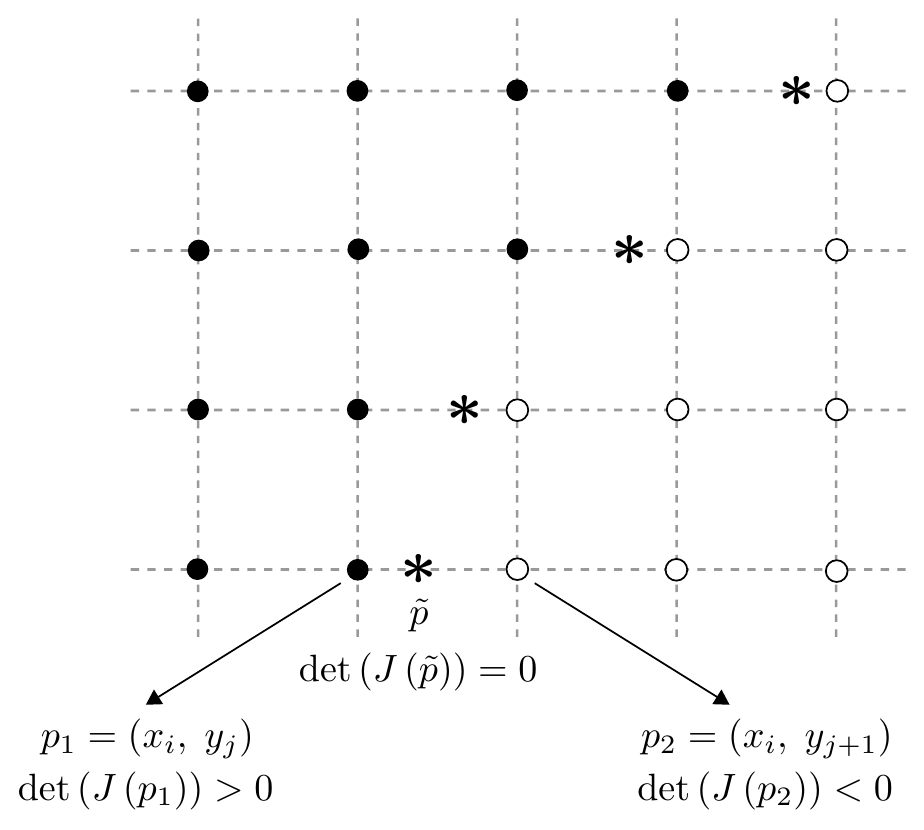}
\caption{Analysis of signs of the determinant of the Jacobian $\bold{J}$: $\bullet$ -- grid points with $\det J>0$; $\circ$ -- grid points with $\det J<0$; $*$ -- points on grid lines
 with $\det J=0$ (critical points).}
\label{fig:signs_critical_curve}
\end{figure}

The determination of the critical curves is performed, in this work, using a numerical continuation method. First, we perform an analysis of the signs of the $\det(\bold{J}(\bold{p}))$ in the domain of interest, using a rectangular grid. Between two neighboring points $ \bold{p_1} $ and $ \bold{p_2} $ in the mesh, in which a change in the signs is identified, the existence of a root $\bold{\tilde{p}}$ given by $\det(\bold{J}(\bold{\tilde{p}})) = 0$ is guaranteed. By traversing the mesh, any of the roots (usually the first that is obtained) can be used as the initial estimate of a continuation method to construct a critical curve. Figure~\ref{fig:signs_critical_curve} illustrates the grid in the domain with different signs for the determinant of the Jacobian matrix. Critical points are represented by $*$.

The continuation method used to construct the critical curves 
is based on  a simple idea: starting with an initial estimate 
$ \bold{\tilde{p}} $, a new critical point (in the mathematical sense)
 can be obtained in the direction defined by the gradient of 
 $\det(\bold{J}(\bold{\tilde{p}}))$, in a distance defined by 
 a sufficiently small step length. In this way, we estimate
 the location of a new critical point. Then it is possible to define 
 an arbitrarily small line segment that joins two points in the 
 vicinity of the estimate obtained for the critical point. Thus, the 
 achievement of the new critical point results from the calculation of 
 the root of the line segment considered. If there is no root in the 
 straight line segment, its size should vary adaptively until a 
 critical point is found. This procedure is repeated until the 
 critical curve is fully constructed.

\subsection{Construction of the bank of solved points}

In the second step, a bank of solved points is constructed. The main purpose of this structure is to store the initial conditions for the inversion process. The bank is composed of solutions for the nonlinear algebraic system $\bold{F}(\bold{p})=\bold{q}$, obtained by a classical Newton method when $ \bold{q} \neq \bold{0} $. This procedure will produce a set of pre-images associated with the respective images in the vicinity of $ \bold{q} = \bold{0} $. This phase is the more expensive step (in terms of computational effort and user interference) of the algorithm. But, as the results of this work will demonstrate, a unique bank of solved points can be used to solve problems for mixtures with different compositions. It is worth mentioning that the procedure adopted here is different to that employed by \cite{bib:malta1996numerical}, when the authors used a set of squares in the image which are traversed in an anti-clockwise direction.

\subsection{Inversion of a point in the image}

\begin{figure}[!htbp]
\centering
\includegraphics[width=0.8\textwidth]{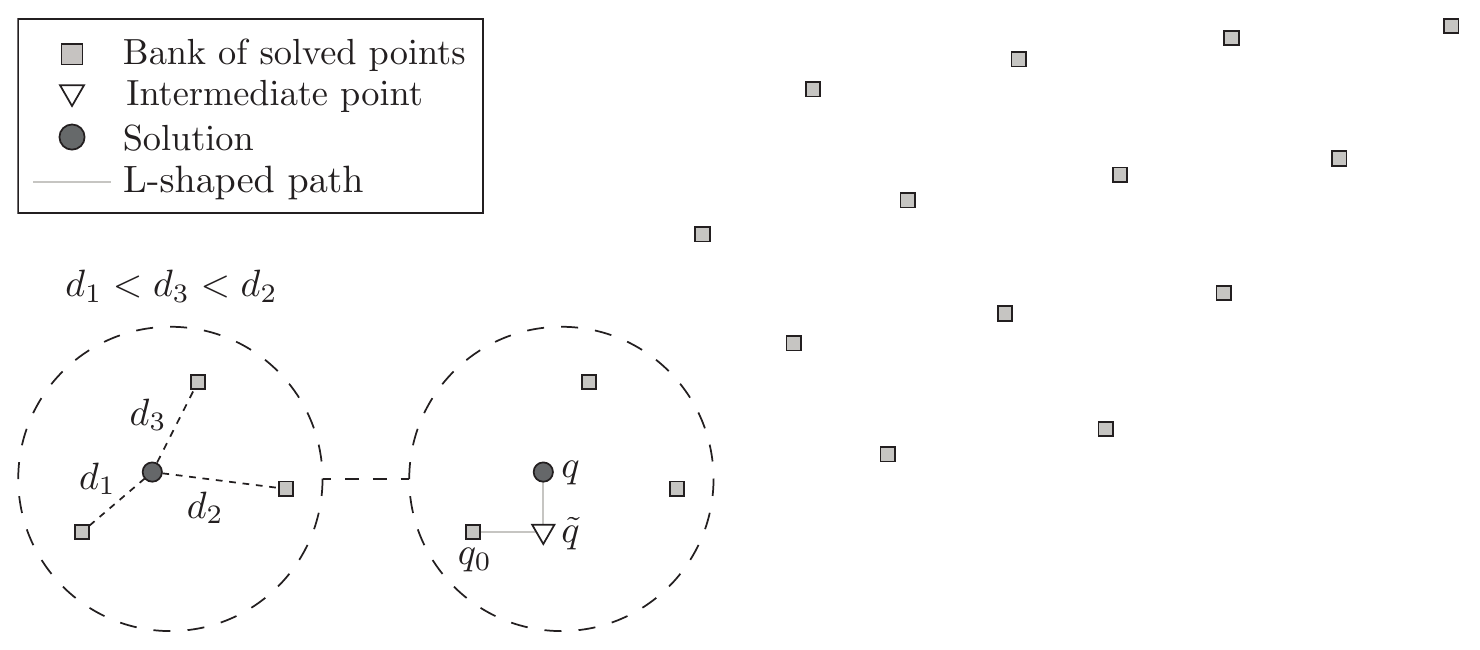}
\caption{The bank of solved points and the L-shaped path in the image.}
\label{fig:inversion_L_path}
\end{figure}

The last step of the methodology is the inversion process \textit{per se}. In this step, a point of the bank of solved points is used as initial condition for the inversion process, conducted by an Euler-Newton predictor-corrector method. The criterion used to select a point in the bank of solved points is the shortest distance to the point to be inverted, in this case, $\bold{q} = \left(0, \; 0\right)$. Figure~\ref{fig:inversion_L_path} represents schematically the bank of solved points (gray squares) and the choice of a particular point in the inversion process. We  intend to invert the point $ \bold{q} $ in the image. We note that there are three points in the vicinity of the point $ \bold{q} $. The algorithm then calculates the distances $d_1$, $d_2$ and $d_3$ and chooses the shortest distance---in this example, $d_1$. The point $\tilde{\bold{q}}$ is an intermediate point in the image (used to produce the ``L''-shaped path).

Starting from the point in the domain corresponding to the one 
selected in the bank of solved points in the image, the inversion 
procedure uses a homotopy technique to obtain the L-shaped path to the 
point $ \bold{p} $, where $\bold{F}(\bold{p})=\left( 0, \; 0 \right)$. 
A predictor-corrector method is iteratively applied to calculate the 
curve implicitly defined by the function deformed using the homotopy 
technique. In the predictor step, a point is calculated using the Euler method, and then the estimated point is corrected using the Newton method. \textcolor{black}{The stopping criterion involves two conditions: when the convergence occurs, the corresponding pre-image related to $ \bold{q} = \left( 0, \; 0 \right) $ is reached. Thus, the algorithm terminates when $ F_{1} \left( p^{\left( \ell \right)} \right)^{2} + F_{2} \left( p^{\left( \ell \right)} \right)^{2} < 10^{-12} $, where $ \ell $ is the iteration counter. Otherwise, the execution stops when a maximum number of iterations is reached.} Detailed information about implementation of Euler-Newton predictor-corrector methods can be seen in \citet[p.~48]{allgower}.

Here, some precautions are necessary: the path of the inversion cannot cross critical curves\footnote[1]{Henceforth, the expression ``critical curve'' refers to the mathematical sense, whereas ``critical point'' represents the thermodynamic critical point. \textcolor{black}{We will use the expression ``thermodynamic critical curve'' to refer to the critical curve in the thermodynamic sense.}}, since this condition implies in a change of the number of pre-images of the problem (in a fold, the number of pre-images varies, +2 or $-2$, \citet{bib:malta1996numerical}). A glimpse of why this happens in nonlinear functions can be seen in Section~\ref{31}. Thus, small paths are preferred, using points of the bank in the vicinities of the solution (automatically evaluated by the computational code, using the Euclidean distance between the point of the bank and the desired solution in the image of $\mathbf{F}$, which is a known value).

\section{Results} \label{sec:results}

In this section we present some numerical results using the proposed approach, for \textcolor{black}{four mixtures}: (i) ethane + methane; (ii) methane + hydrogen sulfide; \textcolor{black}{(iii) methane + ethanol; (iv) cyclohexane + carbon dioxide}. Critical properties and acentric factors for the substances used in this work are presented in Table \ref{tab:dados_mistura}. \textcolor{black}{We are considering only stable critical points in our computations. For instance, critical points found in the mixture methane + hydrogen sulfide were previously addressed by \citet{nichita2010} and the stability analysis was provided by these authors for the same compositions analyzed here. Thus, we can assure that the critical coordinates correspond to stable critical points.}

\begin{table}[!htbp]
\caption{Critical properties and acentric factor for the mixture components}
\centering
{\setlength{\tabulinesep}{1.2mm}
\begin{tabu}{lcccc}
\hline
Component                 & $P_c\ \text{(kPa)}$ & $T_c\ \text{(K)}$ &  $\omega$ & Extracted from \\
\hline
Ethane                    & 4872    & 305.32 &    0.099 & \textcolor{black}{\citet{Dimitrakopoulos}}  \\
\textcolor{black}{Ethanol} & \textcolor{black}{6148}    & \textcolor{black}{513.92} &    \textcolor{black}{0.644} & \textcolor{black}{\citet{bib:stryjek1986}}
\\
Methane                   & 4599    & 190.56 &    0.011 & \textcolor{black}{\citet{Dimitrakopoulos}}  \\
Hydrogen Sulfide          & 9000    & 373.10 &    0.081 & \textcolor{black}{\citet{bib:chapoy2005}}
\\
\textcolor{black}{Cyclohexane} & \textcolor{black}{4075}    & \textcolor{black}{553.64} &    \textcolor{black}{0.208} & \textcolor{black}{\citet{bib:stryjek1986}}
\\
\textcolor{black}{Carbon Dioxide} & \textcolor{black}{7382}    & \textcolor{black}{304.21} &    \textcolor{black}{0.225} & \textcolor{black}{\citet{bib:stryjek1986}}
\\
\hline
\end{tabu}}
\label{tab:dados_mistura}
\end{table}

\subsection{Example 1: mixture ethane (1) + methane (2)}

In order to illustrate the application of the methodology, we calculate the critical point for a mixture containing 90\% of ethane and 10\% of methane, in molar quantities. \textcolor{black}{In this mixture, we employed the Peng-Robinson EOS with van der Waals-I mixing rules and classical combination rules \citep{peng1976new}. We used $k_{12} = 0.0026$, according to \citet{fateen2013}.}

\begin{figure}[!htbp]
\centering
\includegraphics[scale=0.9]{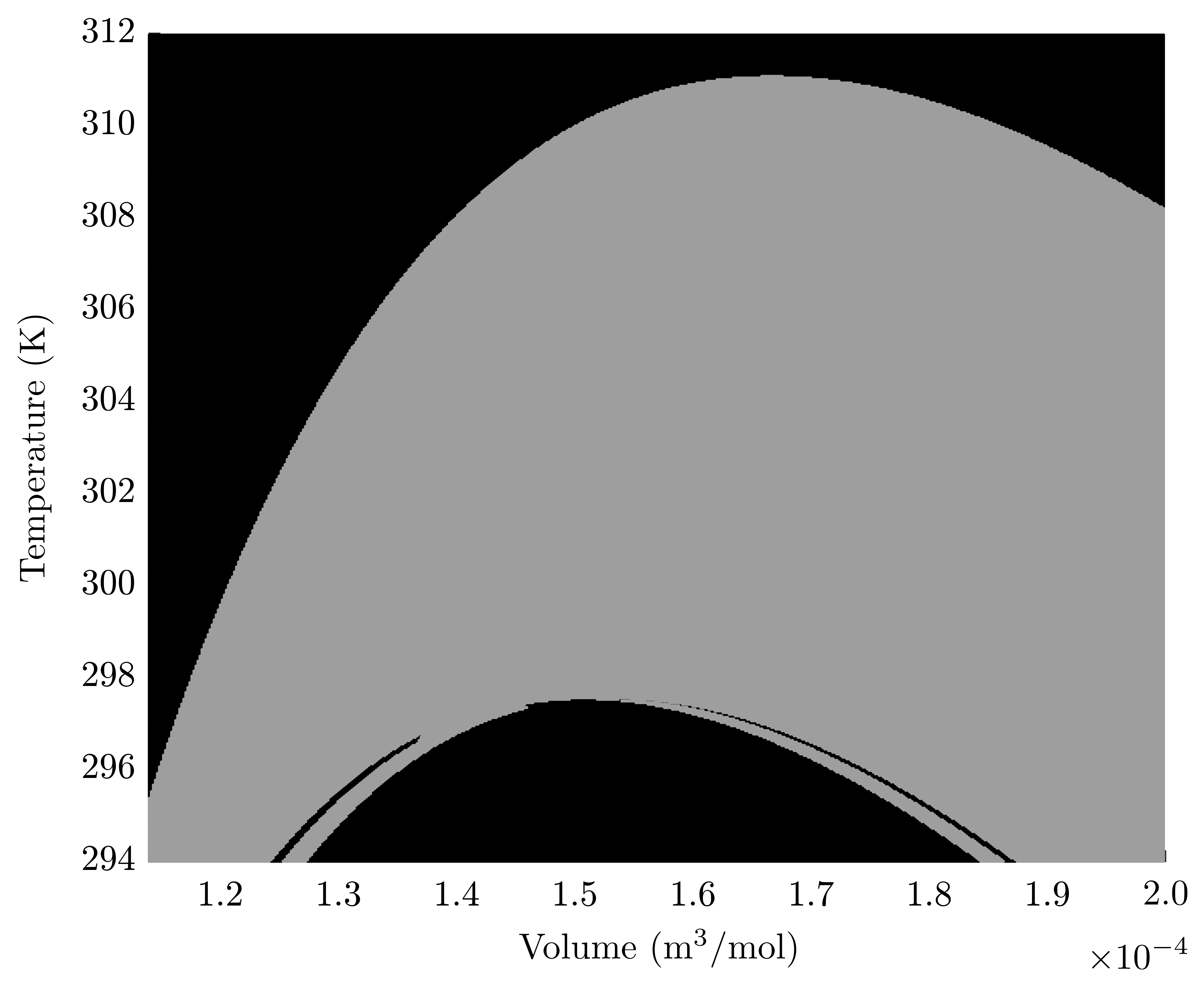}
\caption{Signs of the determinant of the Jacobian matrix for the mixture ethane (1) $+$ methane (2).}
\label{fig:signals}
\end{figure}

As said before, the first computational step is the determination of the critical set of the function. A possible visualization of the critical curve is obtained by plotting the signs of the determinant of the Jacobian matrix with different shades of gray.
 Figure \ref{fig:signals} illustrates the sign of the determinant as function of $T$ and $V$, where the portions in black and in gray have opposite signs. 

\begin{figure}[!htbp]
\centering
\includegraphics[scale=0.9]{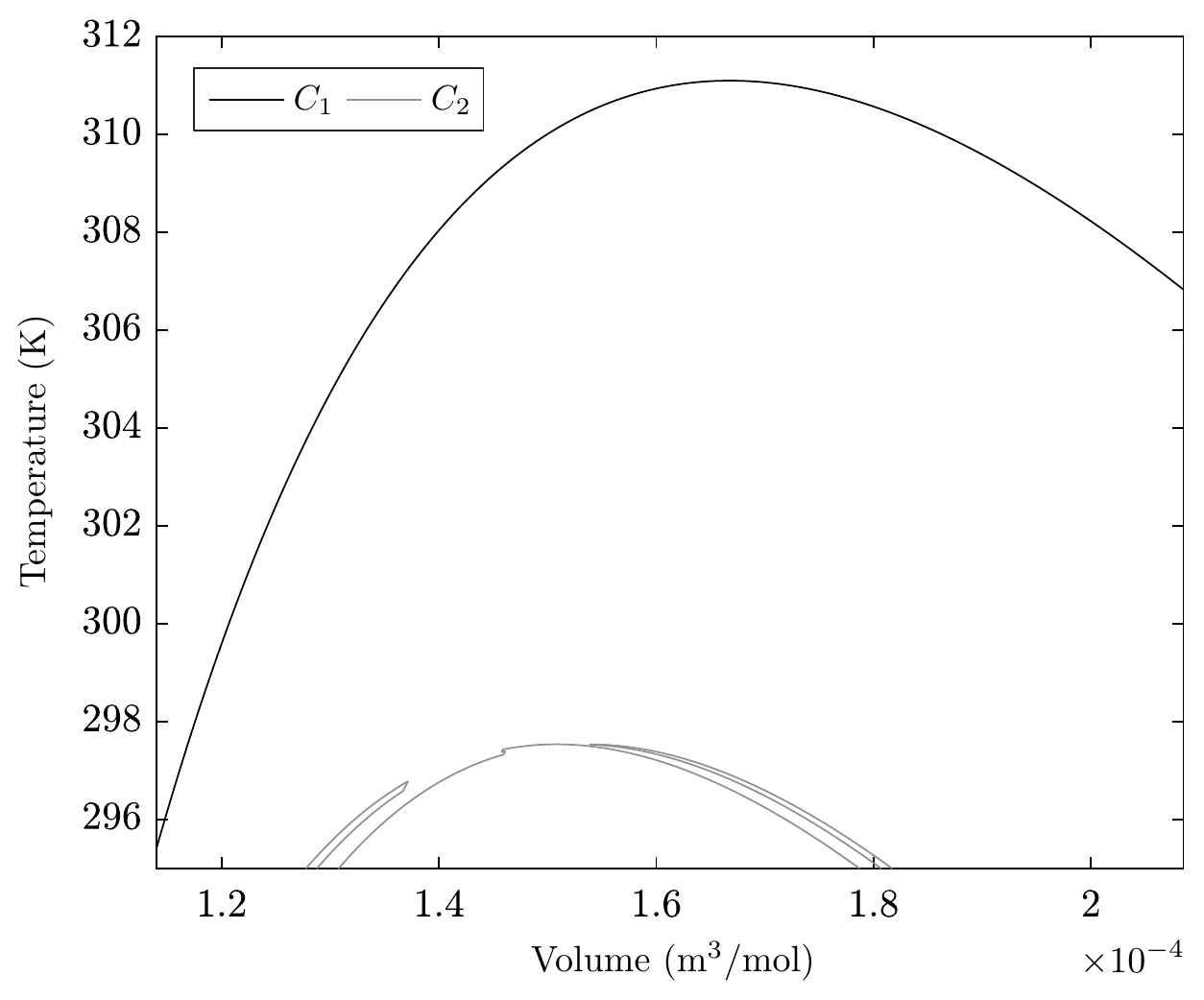}
\caption{Critical curves in the system ethane (1) + methane (2).}
\label{fig:curvas_criticas_problema1}
\end{figure}

Through this representation, it is possible to estimate
 the behavior (and positioning) of the critical curves, since the points that make up each of them are located at the interface between the two shades of gray. The critical curves are detailed in Figure~\ref{fig:curvas_criticas_problema1}. We 
  note a clear correspondence between the images presented in Figure~\ref{fig:signals} and in Figure~\ref{fig:curvas_criticas_problema1}.

\begin{figure}[!htbp]
\centering
\includegraphics[scale=0.9]{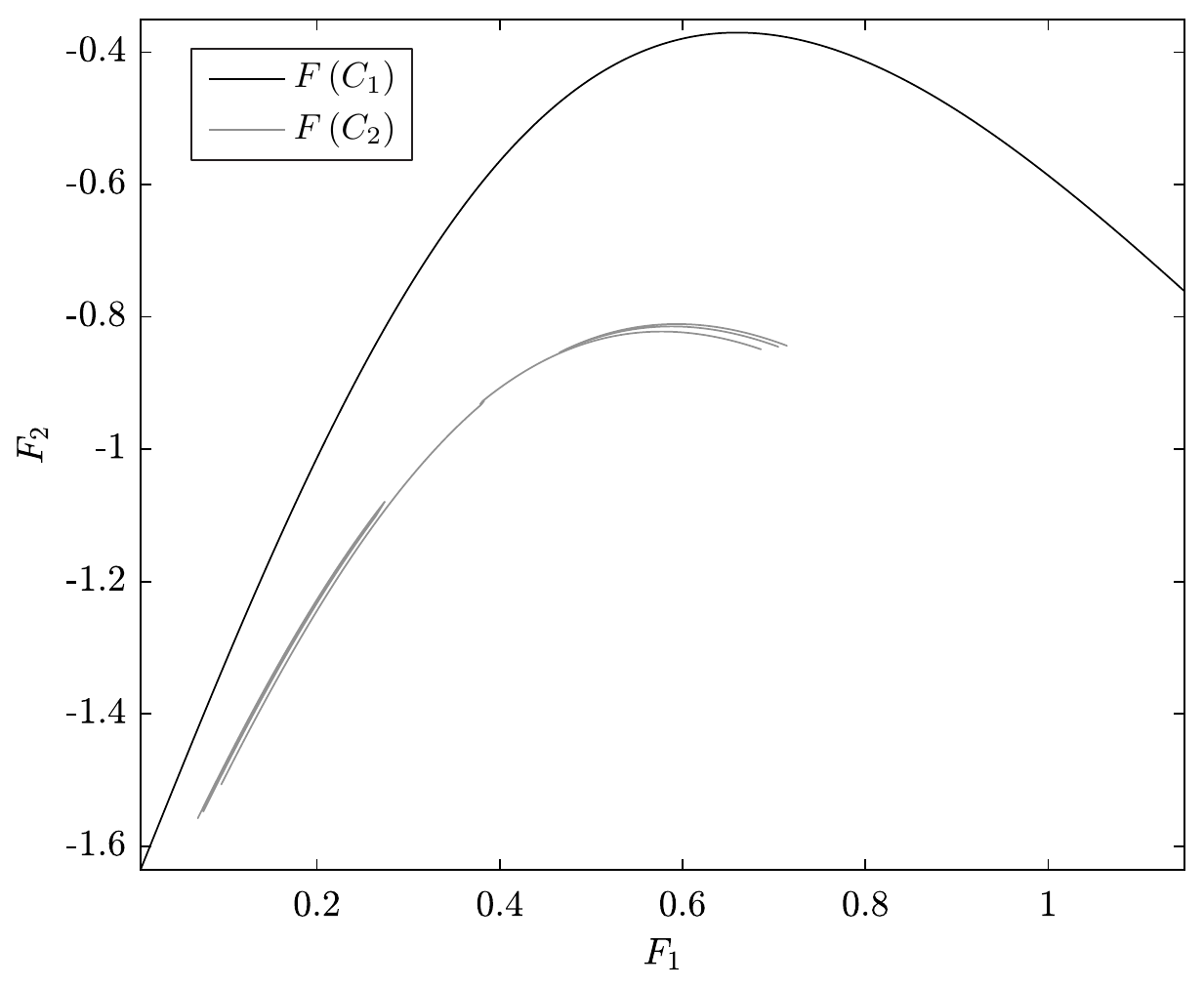}
\caption{Critical images in the system ethane (1) + methane (2).}
\label{fig:imagens_criticas}
\end{figure}

\begin{figure}[!htbp]
\centering
\includegraphics[scale=0.9]{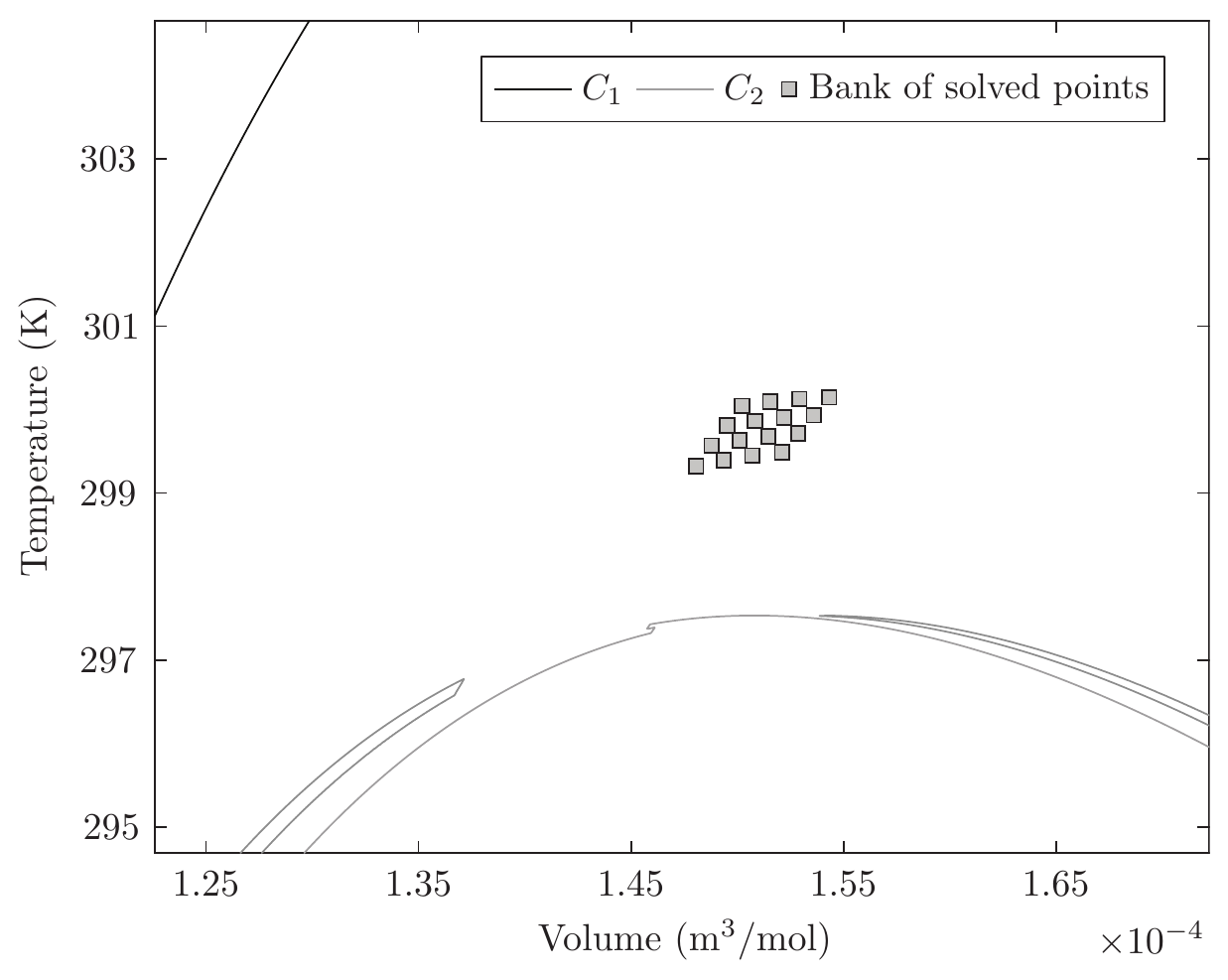}
\caption{Critical curves and the bank of solved points.}
\label{fig:banco_pontos_resolvidos_dominio}
\end{figure}

The next step is to obtain the critical images of the function, by 
evaluating the critical set using the function defined in
Equation~(\ref{final2}), which 
describes the thermodynamic critical point conditions 
by Equation~(\ref{final}), accordingly to 
the approach by \cite{HeidemannKhalil}. The critical image of the 
function is represented in Figure~\ref{fig:imagens_criticas}. The bank 
of solved points (in the domain) is illustrated in 
Figure \ref{fig:banco_pontos_resolvidos_dominio}. In this problem, only one pre-image 
exists.

\begin{figure}[!htbp]
\centering
\includegraphics[scale=0.9]{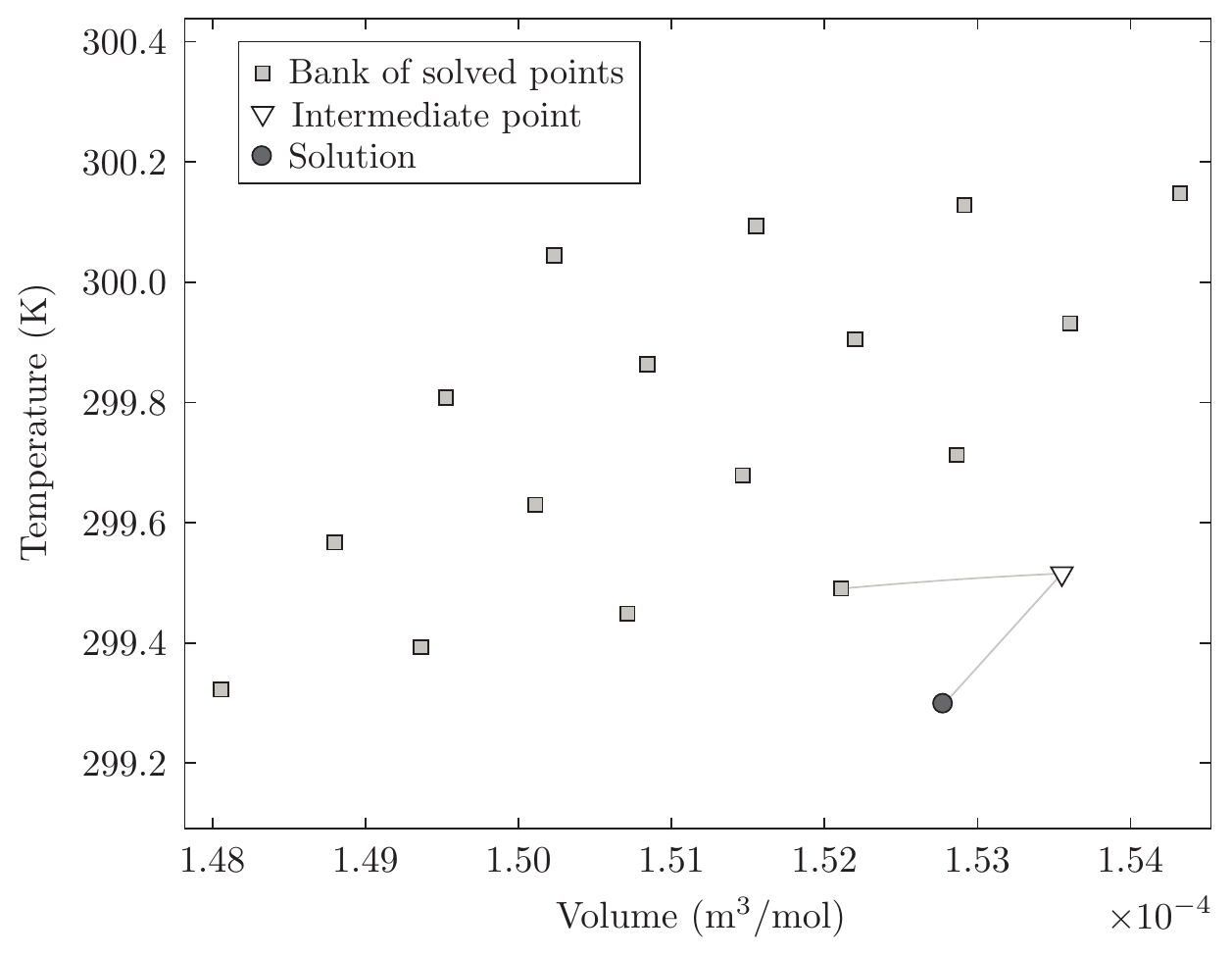}
\caption{Inversion path in the domain for the point $q = (0,0)$.}
\label{fig:inversao_dominio}
\end{figure}

Figure \ref{fig:inversao_dominio} detaches the path of the inversion process in the domain. We can also note that this inversion process maintains the two-step process, corresponding to the L-shaped path in the image, as illustrated in Figure \ref{fig:inversion_L_path}.

The final results obtained by the numerical inversion of this function are presented in Table~\ref{tab:resultado_final_prob1}, which also shows a comparison between these results and the ones produced by the methodology proposed by \citet{Dimitrakopoulos}, that uses a damped Newton's method for temperature, volume and the numbers of moles vector $\bold{\Delta} \bold{n}$. The results are fully compatible, indicating that the methodology is capable to produce accurate results.

\begin{table}[!htbp]
\caption{\textcolor{black}{Critical point in the mixture ethane + methane (90\% of ethane and 10\% of methane) using the numerical inversion of functions (this work) and the damped Newton approach \citep{Dimitrakopoulos}.}}
\centering
{\setlength{\tabulinesep}{1.2mm}
\begin{tabu}{lcc}
\hline
                                   & Inversion of functions           & Damped Newton \\
\hline
$P_c\ \text{(kPa)}$                & 5309.9                            & 5312.0  \\
$T_c\ \text{(K)}$                  & 299.3                             & 299.0   \\
$V_c\ (\text{m}^3/\text{mol})$     & 1.5 $\times$ 10$^{-4}$       & not informed \\
\hline
\end{tabu}}
\label{tab:resultado_final_prob1}
\end{table}

\subsection{Example 2: mixture methane (1) + hydrogen sulfide (2)}

The second mixture tested is composed by methane (1) + hydrogen sulfide (2), also modeled by the Peng-Robinson EOS \textcolor{black}{using van der Waals-I mixing rules} \textcolor{black}{\citep{peng1976new}}. For some compositions this mixture exhibits more than one critical point \citep{stradi}. A detailed discussion of the thermodynamic behavior of the mixture methane + hydrogen sulfide was presented by \cite{lange2016}.

We must emphasize that \cite{stradi} employed the Soave-Redlich-Kwong EOS \textcolor{black}{\citep{bib:srk}} in their computations, while we are using the Peng-Robinson EOS. 
Thus, some differences in the results may be consequences of the 
different thermodynamic models used. We employed $k_{12} = 0.08$ for 
the binary interaction parameter for the pair methane-hydrogen sulfide---the same value employed by \cite{stradi}, although this value was 
obtained to the Soave-Redlich-Kwong equation of state; on the other 
hand, the values for Peng-Robinson and Soave-Redlich-Kwong equations 
are similar \citep{fateen2013}. \cite{nichita2010} also used $k_{12} = 0.08$ for the Peng-Robinson EOS in critical point calculations. Furthermore, \cite{carroll1995} obtained the optimum value for the binary interaction parameter using the Peng-Robinson EOS for the pair methane + hydrogen sulfide equal to $k_{12} = 0.083$. Nevertheless, the main goal here is to verify the robustness and accuracy of the methodology, but not regarding the aspects of the model; in this sense we are not particularly aimed at  checking
 the differences of calculated and experimental values.

\begin{figure}[!htbp]
\centering
\includegraphics[scale=0.9]{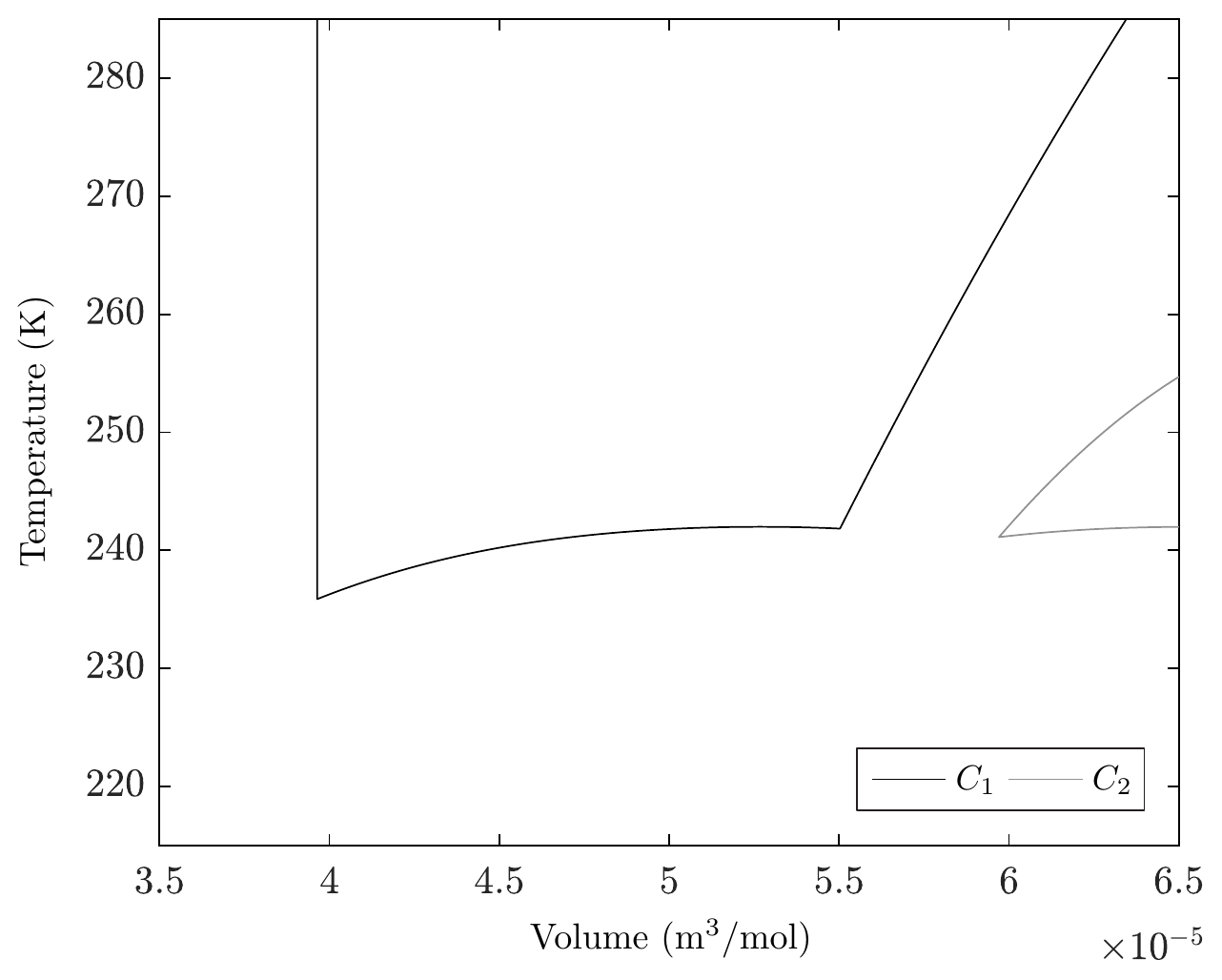}
\caption{Critical curves in the system methane (51\%) + hydrogen sulfide (49\%).}
\label{fig:curvas_criticas_problema2}
\end{figure}

\begin{figure}[!htbp]
\centering
\includegraphics[scale=0.9]{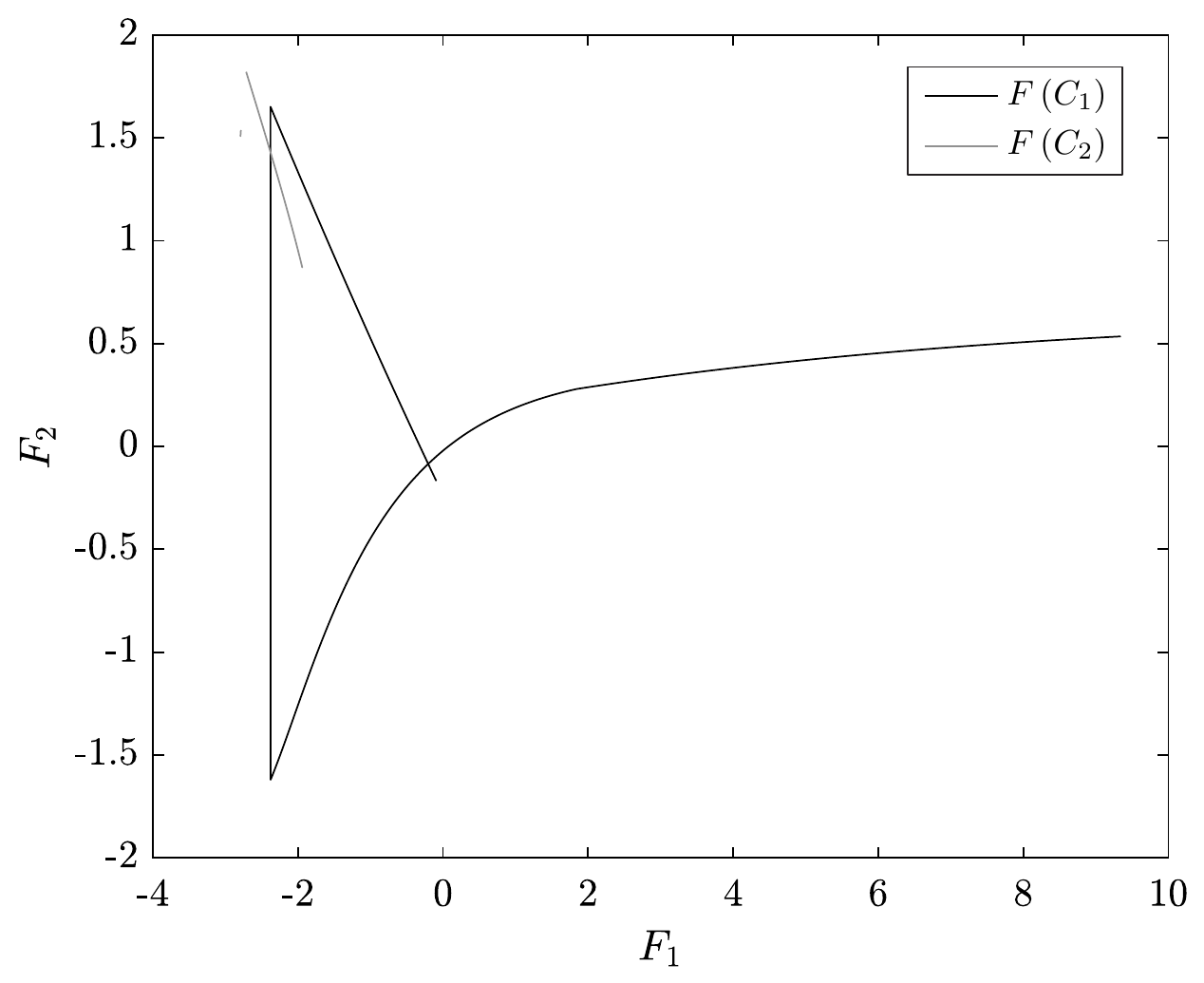}
\caption{Critical images in the system methane (51\%) + hydrogen sulfide (49\%).}
\label{fig:imagens_criticas_problema2}
\end{figure}

The mathematical critical curves for this mixture, with 51\% of methane and 49\% of hydrogen sulfide, are depicted in Figure~\ref{fig:curvas_criticas_problema2}. In turn, Figure \ref{fig:imagens_criticas_problema2} contains the critical images for this problem.

\begin{figure}[!htbp]
\centering
\includegraphics[scale=0.9]{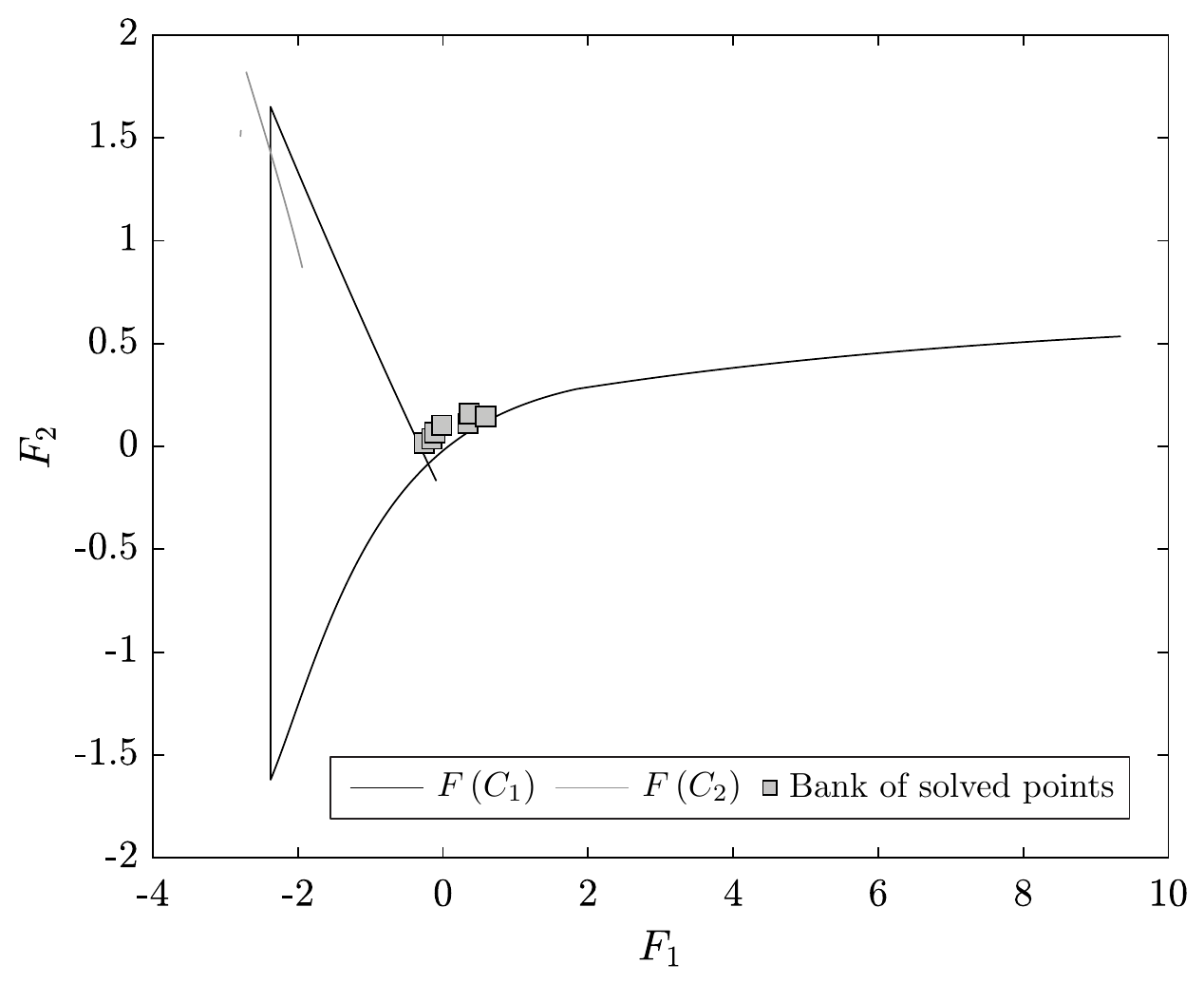}
\caption{Bank of solved points in the image for the system methane (51\%) + hydrogen sulfide (49\%).}
\label{fig:banco_imagem_problema2}
\end{figure}

\begin{figure}[!htbp]
\centering
\includegraphics[scale=0.9]{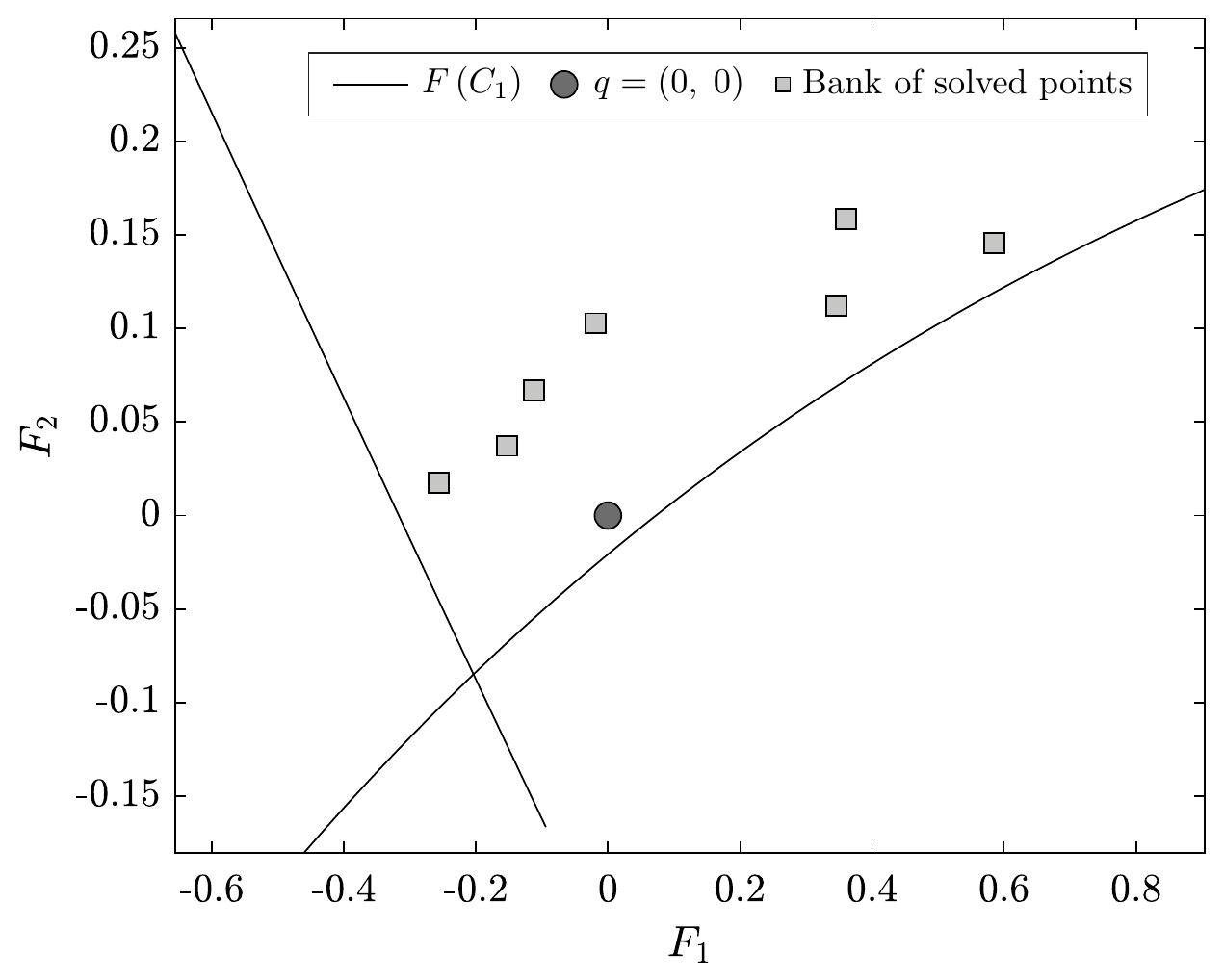}
\caption{Bank of solved points in the image for the system methane (51\%) + hydrogen sulfide (49\%) (amplification).}
\label{fig:banco_imagem_problema2_amp}
\end{figure}

Again, a bank of solved points is necessary, as pointed out previously. The bank of solved points is presented in Figure \ref{fig:banco_imagem_problema2}. Since the points are very close, an amplification is 
provided in Figure \ref{fig:banco_imagem_problema2_amp}.

\begin{figure}[!htbp]
\centering
\includegraphics[scale=0.9]{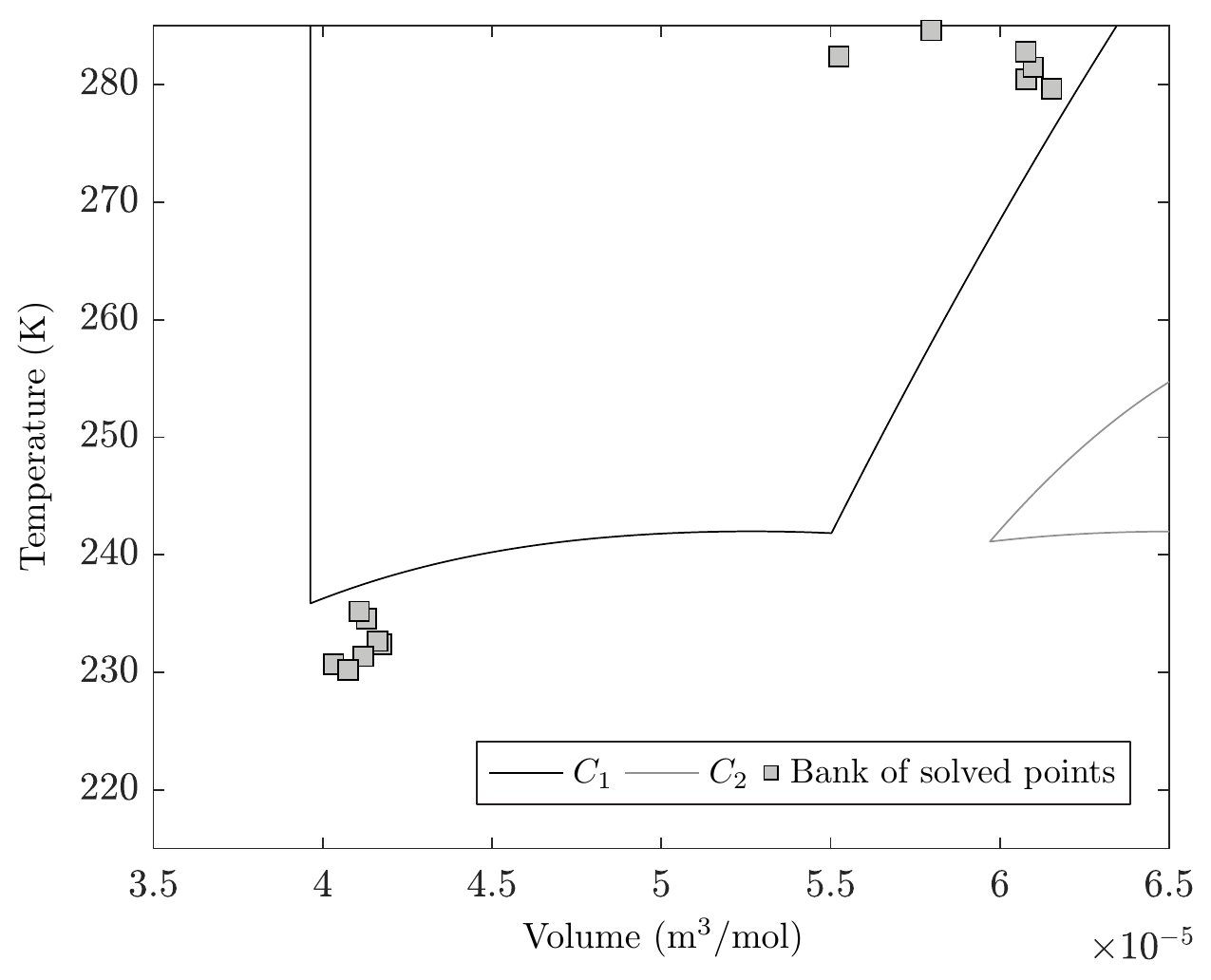}
\caption{Bank of solved points in the domain for the system methane (51\%) + hydrogen sulfide (49\%).}
\label{fig:banco_problema2_dominio}
\end{figure}

Figure~\ref{fig:banco_problema2_dominio} contains the bank of solved points in the domain (i.e., the pre-images produced with good initial estimates and using a Newton algorithm). Clearly, there are
two group of pre-images, since this mixture shows two critical points (in the thermodynamic sense) for this composition. Also, note that the sets of pre-images are concentrated in distinct regions in the domain, bounded by the critical curve. Since the L-shaped path produced by the continuation method must not cross a critical curve (due to solution degeneracy), such positioning of the points is desirable, from the algorithmic point of view.

\begin{figure}[!htbp]
\centering
\includegraphics[scale=0.9]{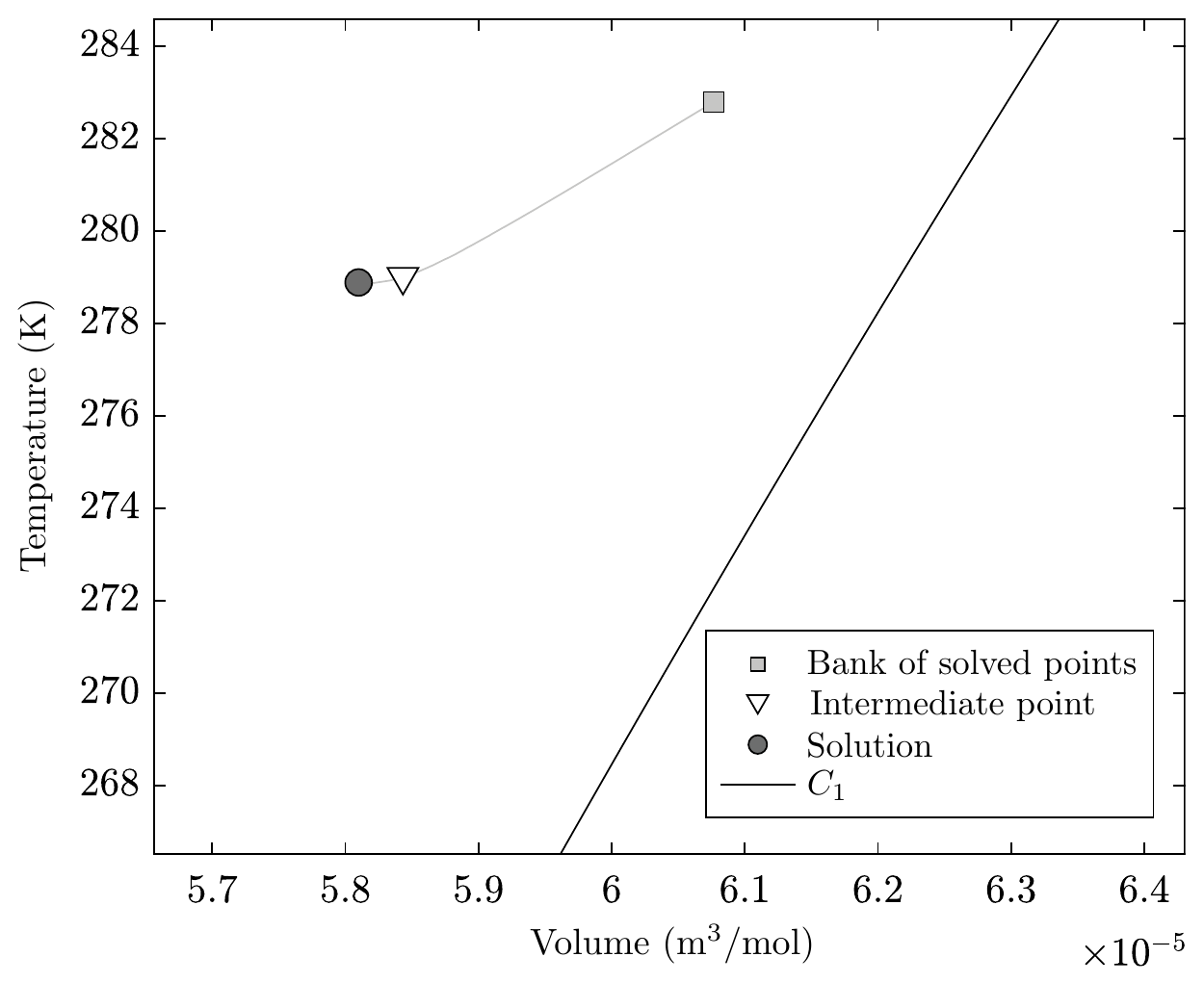}
\caption{Inversion path in the domain for the critical point 1 in the system methane (51\%) + hydrogen sulfide (49\%).}
\label{fig:inversao_problema2_ponto1}
\end{figure}

\begin{figure}[!htbp]
\centering
\includegraphics[scale=0.9]{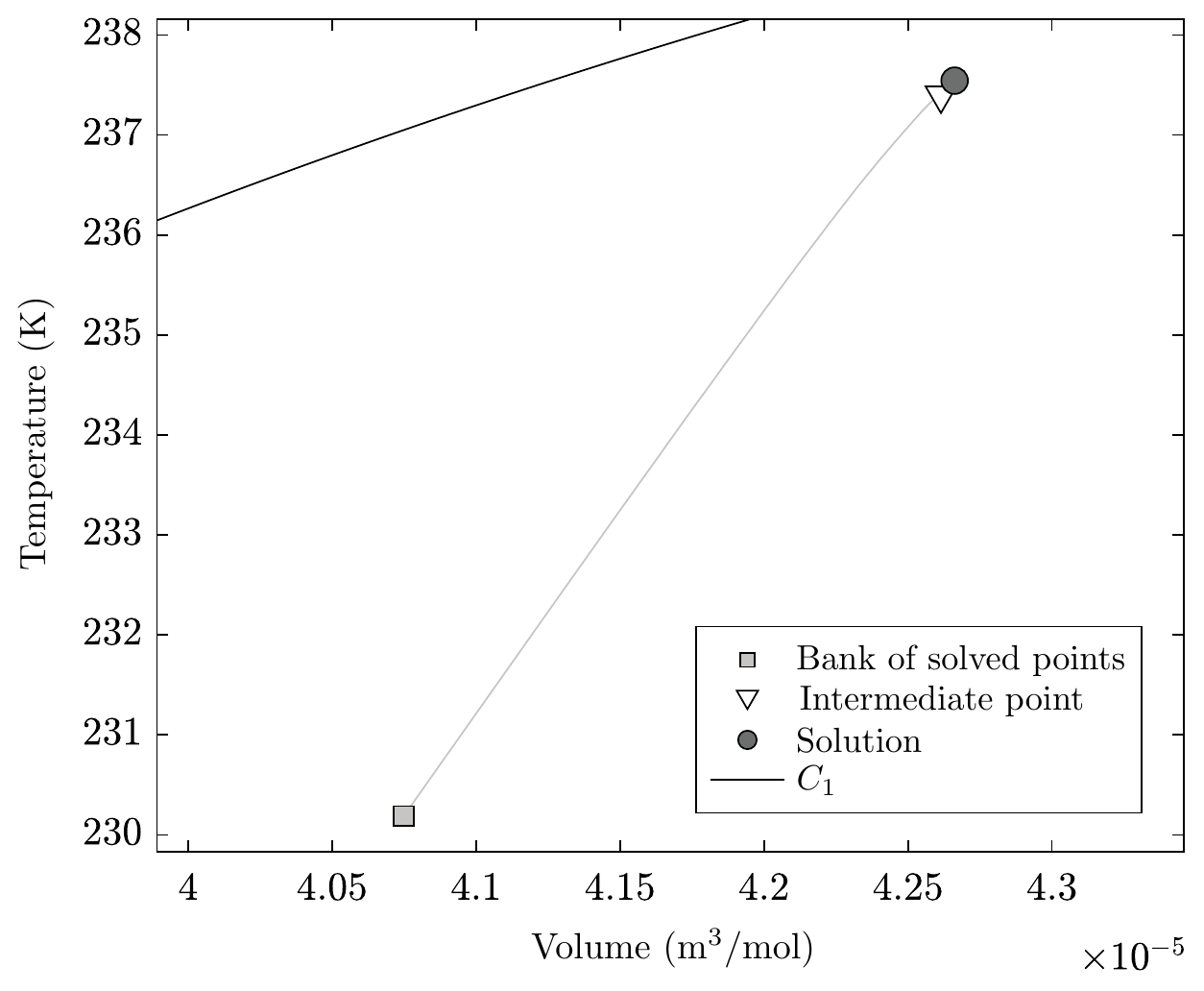}
\caption{Inversion path in the domain for the critical point 2 in the system methane (51\%) + hydrogen sulfide (49\%).}
\label{fig:inversao_problema2_ponto2}
\end{figure}

Finally, the inversion paths for the critical points 1 and 2 are presented in Figures \ref{fig:inversao_problema2_ponto1} and \ref{fig:inversao_problema2_ponto2}, respectively.

Table~\ref{tab:resultado_final_prob2} presents the computed critical points in the mixture methane + hydrogen sulfide by the methodology
described here (numerical inversion of functions), by
using the classical approach of \citet{HeidemannKhalil} (using a double-loop structure with two Newton solvers for temperature and molar volume) and also by employing a stochastic optimization algorithm 
\citep{nichita2010}. \textcolor{black}{The first critical point (higher temperature) corresponds to a liquid-vapor critical point, whereas the second one (lower temperature) is a liquid-liquid critical point. Both critical points belong to the same branch of the thermodynamic critical curve. This mixture is classified as Type III in the classification of van Konynenburg and Scott \citep{lange2016}.}

We observe that the values of the critical coordinates---calculated using absolutely different methodologies---are virtually the same; small deviations are probably consequence of different parameters for pure substance and/or parameters of Peng-Robinson EOS.
Also, we note that for the first two approaches, the values of the residues $F_1$ and $F_2$ are close to zero, with only slight variations, indicating that the critical conditions represented by Equations~(\ref{sistema_final}) are satisfied. 




\begin{table}[!ht]
\centering
\caption{\textcolor{black}{Critical points in the mixture methane + hydrogen sulfide, with two different compositions. Comparison of methods: numerical inversion of functions (this work); H\&K (\citet{HeidemannKhalil}); N\&G (\cite{nichita2010}).}}
\label{tab:resultado_final_prob2}
\resizebox{\textwidth}{!}{%
{\setlength{\tabulinesep}{1.3mm}
\begin{tabu}{ccccccc}
\hline\hline
Method & Composition & $ T_{c} \left( \mathrm{K} \right) $ & $ V_{c} \left( \mathrm{m^{3} / mol} \right) $ & $ P_{c} \left( \mathrm{kPa} \right) $ & $ F_{1} \left( V_{c}, \; T_{c} \right) $ & $ F_{2} \left( V_{c}, \; T_{c} \right) $ \\ \hline
\multirow{4}{*}{This work} & \multirow{2}{*}{51\%/49\%} & $278.89$ & $5.81\times 10^{-5}$ & $14213.85$ & $-8.92\times 10^{-10}$ & $-1.79\times 10^{-9}$  \\
 &  & $237.54$ & $4.27\times 10^{-5}$ & $16056.88$ & $6.60\times 10^{-9}$ & $-1.78\times 10^{-8}$ \\
 & \multirow{2}{*}{52\%/48\%} & $273.17$ & $5.59\times 10^{-5}$ & $14193.38$ & $-2.91\times 10^{-8}$ & $-1.89\times 10^{-7}$ \\
 &  & $245.43$ & $4.52\times 10^{-5}$ & $14776.31$ & $3.62\times 10^{-10}$ & $-1.64\times 10^{-7}$ \\ \hline
\multirow{4}{*}{H\&K} & \multirow{2}{*}{51\%/49\%} & $ 278.89$ & $5.81\times 10^{-5}$ & $14213.85$ & $5.57\times 10^{-10}$ & $-2.25\times 10^{-8}$ \\
 &  & $ 237.54$ & $4.27\times 10^{-5}$ & $16056.88$ & $6.38\times 10^{-10}$ & $-1.66\times 10^{-10}$ \\
 & \multirow{2}{*}{52\%/48\%} & $ 273.17$ & $5.59\times 10^{-5}$ & $14193.38$ & $2.26\times 10^{-9}$ & $1.76\times 10^{-8}$ \\
 &  & $ 245.43$ & $4.52\times 10^{-5}$ & $14776.31$ & $-9.53\times 10^{-9}$ & $-6.80\times 10^{-9}$ \\ \hline
\multirow{4}{*}{N\&G} & \multirow{2}{*}{51\%/49\%} & $ 277.80$ & --- & $14330$ & --- & --- \\
 &  & $ 240.26$ & --- & $15889$ & --- & --- \\
 & \multirow{2}{*}{52\%/48\%} & $ 271.40$ & --- & $14316$ & --- & --- \\
 &  & $ 248.56$ & --- & $14783$ & --- & --- \\ \hline\hline
\end{tabu}}}
\end{table}




The results presented in Table~\ref{tab:resultado_final_prob2}  can 
motivate the questioning of why to use such an intricate algorithm---such 
as the inversion of functions in the plane---if the method by 
\citet{HeidemannKhalil} is able to obtain the coordinates with the 
same precision.
A compact answer for this question may be: the technique of \citet{HeidemannKhalil} demands good initial estimates to produce the two critical points. The justification for this answer is supported by the insightful analysis of the performance of the \citet{HeidemannKhalil} algorithm in the calculation of critical points for the mixture methane + hydrogen sulfide presented by \citet{parajara2017}. These authors constructed a two-dimensional diagram---\textit{basin of attraction}---for the critical point calculation in the referred mixture using a double-loop structure (with nested Newton methods) and the \citet{HeidemannKhalil} algorithm. A basin of attraction represents---using a color scheme---the convergence pattern for different initial estimates. The results obtained by \citet{parajara2017} indicated that a good quantity of initial estimates do not produce convergent results, due to failure in the inversion of the Jacobian matrix during the execution of the Newton method, for instance. Furthermore, a small portion of the diagram is occupied by the critical point 1 (in other words, critical point 2 is not easily obtained by this particular implementation of the Newton method). This situation indicates that the task of obtaining multiple critical points can be hard to accomplish by using a strategy based on Newton method.

Therefore, there is no ``best algorithm'' to approach this kind of problem: all algorithms show advantages and drawbacks. Newton based techniques tend to be more rapid (in terms of time of computation); but the existence of unfavorable basins of attraction introduces difficulties to obtain several critical points. On the other hand, the inversion of functions from the plane to the plane demands high user-interference in the initial steps, but permits a good interpretation of the persistence of the multiple critical points, combined to a robust and accurate methodology in the search for the critical coordinates.


As discussed in a previous section, the construction of the bank of solved points is the more expensive step in the methodology. On the other hand, we can use a single bank of solved points to obtain the solutions for different compositions. This strategy was also presented in a double azeotropy problem \citep{GuedesPlattMouraNeto} and in a double retrograde vaporization problem \citep{industrial2018}. In order to illustrate this feature, we calculate the critical points for the mixture using 52\% of methane and 48\% of hydrogen sulfide, using the bank of solved points obtained for a different composition.

Table~\ref{tab:resultado_final_prob2} compares the critical coordinates obtained by all three methods.  The residues of the functions represented by Equation~(\ref{sistema_final}) are also shown.  We note that the results are, again, 
virtually the same. The differences in the residues can only be verified with a high number of decimal places.
Thus, we verify that a single bank of solved points can be used to obtain the critical coordinates for different compositions. This is a desirable feature of the methodology, since 
the construction of the bank of solved points is the more expensive step of the algorithm.



\begin{figure}[!htbp]
\centering
\includegraphics[scale=0.9]{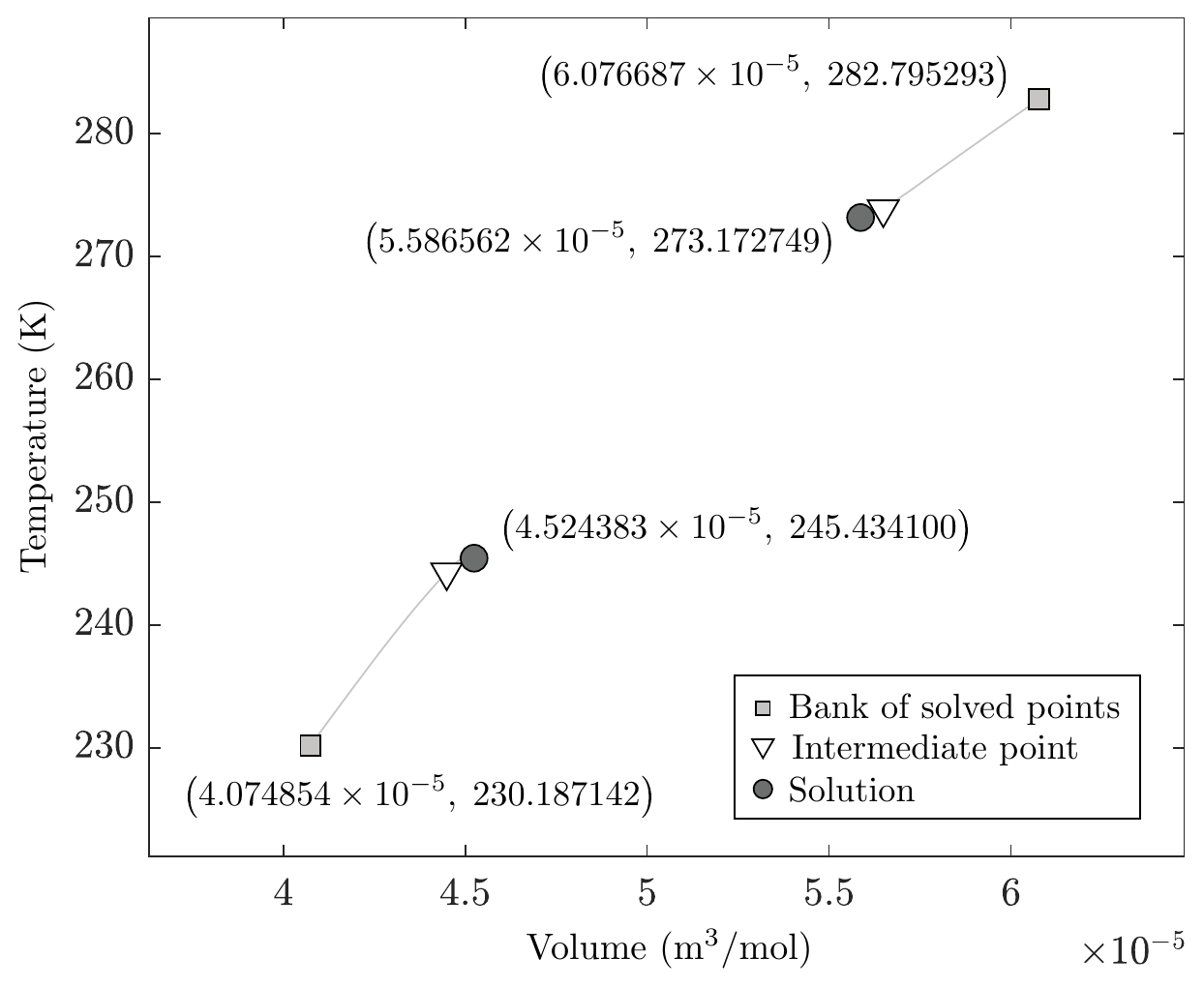}
\caption{Inversion path in the domain for the critical points in the system methane (52\%) + hydrogen sulfide (48\%).}
\label{fig:inversao_problema2_52_48}
\end{figure}

The L-shaped path obtained in the inversion procedure for this situation is presented in Figure~\ref{fig:inversao_problema2_52_48}. The critical curves are not displayed in this figure, due to the scale of the axes. Once again, the paths in the domain are consequences of the L-shaped path in the image.

\subsection{\textcolor{black}{Example 3: mixture methane (1) + ethanol (2)}}

\textcolor{black}{The third mixture analyzed is composed by methane (1) + ethanol (2). Unlike the previous cases, the Stryjek-Vera modification \citep{bib:stryjek1986} of the Peng-Robinson cubic EOS, together with the Wong-Sandler mixing rules \citep{bib:wong1992} and NRTL activity coefficient model \citep{bib:renon1968} are employed to model this mixture. This modification is motivated by the need to verify the robustness of the proposed technique. \citet{bib:castier1997} present a rich analysis about the studied mixture, in addition to providing results about the thermodynamic critical curve, which serves as a benchmark to the results obtained here.}

\textcolor{black}{Acording to \citet{bib:orbey1995}, the binary interaction parameter for this mixture can be adopted as $ k_{ij} = 0 $, without considerable loss of accuracy. In the case of the characteristic pure compound parameter related to the Stryjek-Vera modification of the cubic EOS, \citet{bib:stryjek1986} reports $ \kappa_{1} = \lbrace -0.00159, \; -0.03374 \rbrace $ for methane and ethanol, respectively. For the Wong-Sandler mixing rules, $ \alpha = 0.9 $ and the normalized energies of interaction $ g_{12} / R \left( \mathrm{K} \right) = 165.8 $ and $ g_{21} / R \left( \mathrm{K} \right) = 238.4 $ are adopted in order to calculate de NRTL activity coefficient model \citep{bib:castier1997}.}

\begin{figure}[!htbp]
\centering
\includegraphics[scale=0.9]{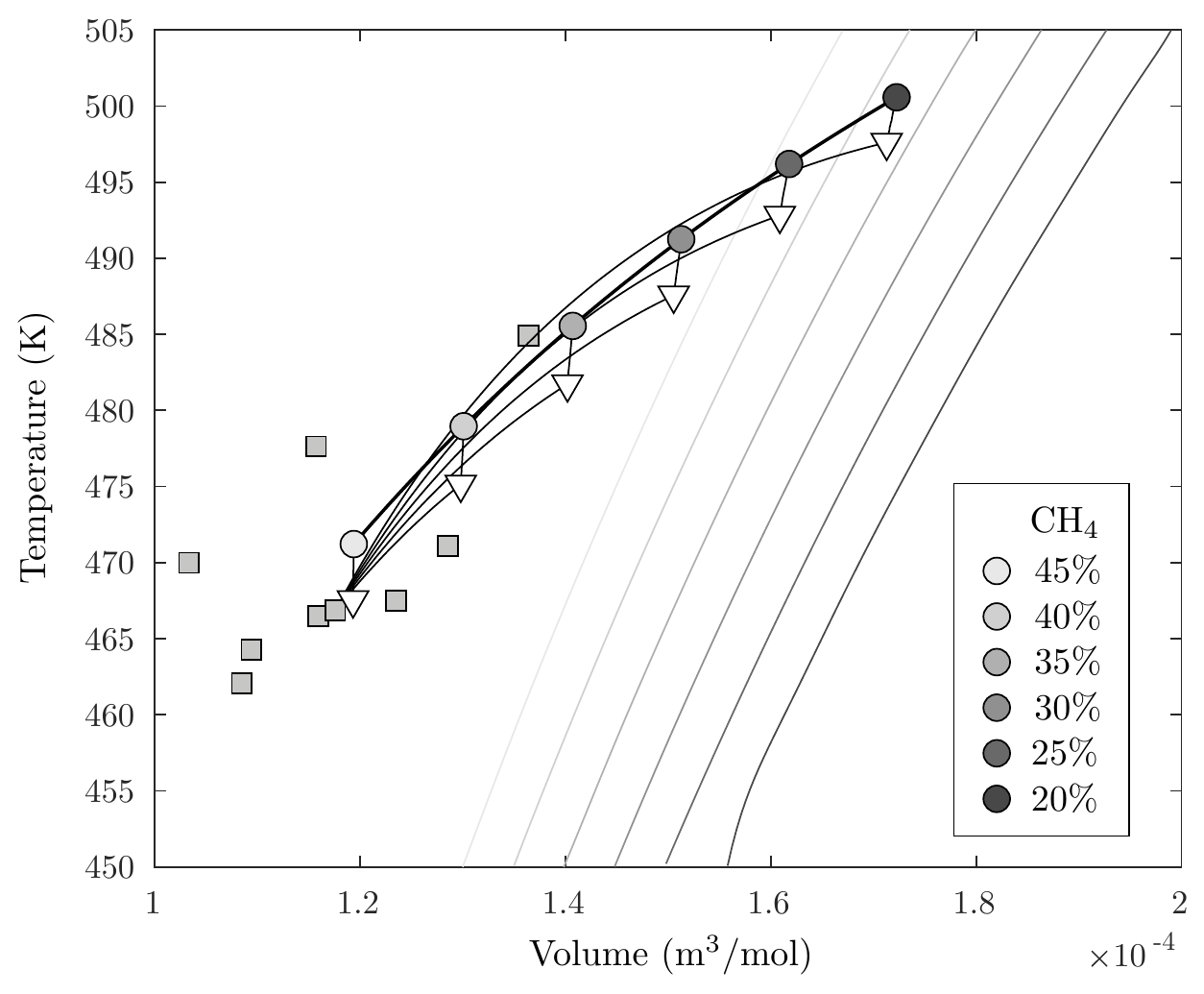}
\caption{\textcolor{black}{Set of results for the inversion procedure and calculation of pre-images of $ q = \left( 0, \; 0 \right) $ of the system methane + ethanol under different compositions. The thicker black line represents a part of the thermodynamic critical curve, fitted using the critical points obtained. The curves of different colors, referring to the $ \mathrm{CH_{4}} $ molar fractions, represent the mathematical critical curves.}}
\label{fig:inversao_variacao_fracao_molar}
\end{figure}

\textcolor{black}{Figure~\ref{fig:inversao_variacao_fracao_molar} shows the results of the application of the method of inversion of functions from the plane to the plane in obtaining the thermodynamic critical points for different concentrations of the components of the mixture. The colors referring to the methane molar fractions (see figure legend) identify the critical points calculated in each of the system configurations and correspond to the mathematical critical curves, that is, the same molar fractions were used to obtain the mathematical critical curve of the same color.}

\textcolor{black}{An important point to note is that all critical points were calculated using the same bank of solved points, which was generated with the composition set at 20\% methane and 80\% ethanol. This fact corroborates the robustness of the method, capable of obtaining satisfactory results even starting from \enquote{inappropriate} initial estimates. In addition, it favors the gain related to the computational effort, given that the generation of the bank of solved points is the most expensive step of the method, from the point of view of the user's interference, allowing the reuse of the same bank in the resolution of correlated problems.}

\textcolor{black}{In all L-shaped paths obtained by the inversion processes shown in Figure~\ref{fig:inversao_variacao_fracao_molar}, the continuation method starts from the same initial estimate until reaching the critical points. This is because, in fact, the initial estimate is chosen by assessing the distance between each of the points in the bank of solved points in the image and the point to be inverted, $ \bold{q} = \left( 0, \; 0 \right) $. Thus, regardless of the concentration of each substance in the mixture, the distance remains the same. Another important point to note is the fact that L-shaped paths, which reach a thermodynamic critical point of a given color, never cross the mathematical critical curve of the same color. As much as the bank of solved points is the same, the resolution of the problems is independent, and this would cause the degeneration of solutions.}

\textcolor{black}{Finally, Figure~\ref{fig:inversao_variacao_fracao_molar} shows a portion of the thermodynamic critical curve, represented by the projection of the calculated discrete thermodynamic critical points and illustrated with a thicker black curve. Shown in the same axes, the difference in relation to the mathematical critical curve is clear. The critical points calculated are shown in Table~\ref{tab:resultados_mistura_3}. By visual inspection, they all coincide with the thermodynamic critical curve calculated by \citet{bib:castier1997}. In addition, the results obtained agree with the values calculated through the method of Heidemann \& Khalil \citep{HeidemannKhalil} for at least six decimal places. Therefore, for the sake of brevity, the values will not be presented in this analysis.}

\begin{table}[!htbp]
\centering
\caption{\textcolor{black}{Critical points in the mixture methane + ethanol obtained through the method of inversion of functions from the plane to the plane for different compositions.}}
\label{tab:resultados_mistura_3}
{\setlength{\tabulinesep}{1.2mm}
\begin{tabu}{cccc} \hline
$ \mathrm{CH}_{4} \left( \% \right) $ & $ P_{c} \left( \mathrm{kPa} \right) $ & $ T_{c} \left( \mathrm{K} \right) $ & $ V_{c} \left( \mathrm{m^{3} / mol} \right) $ \\ \hline
45 & 18407.46 & 471.22 & $ 1.19 \times 10^{-4} $ \\
40 & 16116.52 & 478.96 & $ 1.30 \times 10^{-4} $ \\
35 & 14197.5 & 485.55 & $ 1.41 \times 10^{-4} $ \\
30 & 12560.16 & 491.25 & $ 1.51 \times 10^{-4} $ \\
25 & 11139.49 & 496.19 & $ 1.62 \times 10^{-4} $ \\
20 & 9898.34 & 500.58 & $ 1.72 \times 10^{-4} $ \\ \hline
\end{tabu}}
\end{table}

\subsection{\textcolor{black}{Example 4: mixture cyclohexane (1) + carbon dioxide (2)}}

\textcolor{black}{The last mixture analyzed is composed by cyclohexane (1) + carbon dioxide (2). In terms of molar fractions, the mixture contains 60\% cyclohexane and 40\% carbon dioxide. The Peng-Robinson EOS with van der Waals-I mixing rules and classical combination rules were employed \citep{peng1976new}. Under the same conditions, the thermodynamic critical curve of the mixture has already been presented by \citet{bib:arce2007}. The same authors indicate that the binary interaction parameter for this mixture is $ k_{ij} = 0.0627 $. The other parameters used in the modeling, referring to pure compounds, are listed in Table~\ref{tab:dados_mistura}.}

\begin{figure}[!ht]
\centering
\includegraphics[scale=0.9]{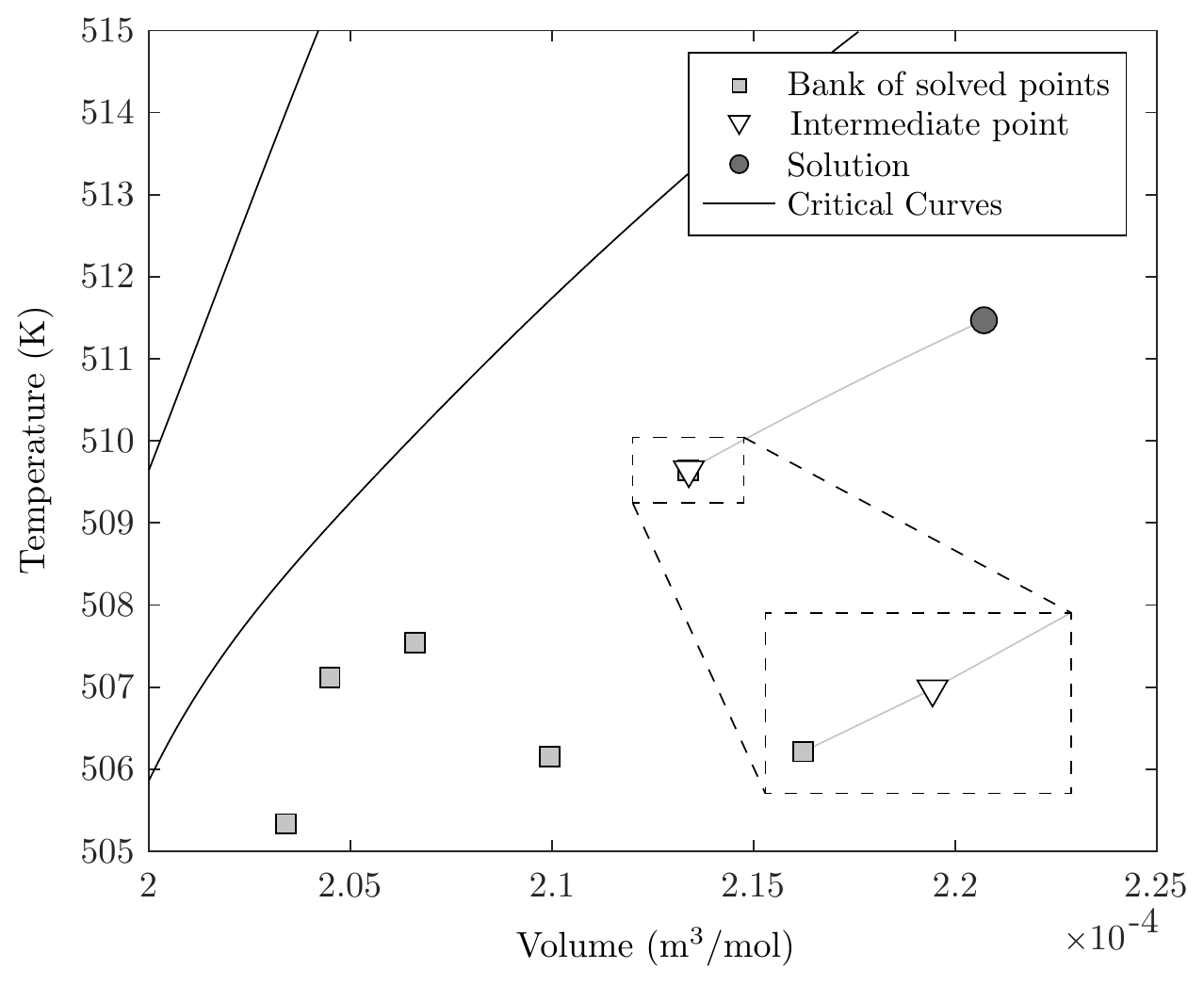}
\caption{\textcolor{black}{Calculation of the L-shaped path until obtaining the pre-image of $ q = \left( 0, \; 0 \right) $ for the system composed by 60\% cyclohexane and 40\% carbon dioxide.}}
\label{fig:cyclohexane_carbon_dioxide}
\end{figure}

\textcolor{black}{Figure~\ref{fig:cyclohexane_carbon_dioxide} shows the L-shaped path, continued from the point of the bank of solved points in the domain to the pre-image of $ \bold{q} = \left( 0, \; 0 \right) $, the critical point of the mixture. The pre-image achieved was $ \bold{p} = \left( 2.21 \times 10^{-4}, \; 511.47 \right) $, with $ \bold{F} \left( \bold{p} \right) = \left( 2.47 \times 10^{-7}, \; -1.59 \times 10^{-8} \right) $. The corresponding critical pressure is $ 8884.52 $ kPa. The critical point calculated by the proposed methodology is found on the thermodynamic critical curve approximated by \citet{bib:arce2007}. The results obtained were compared with those obtained by the method of Heidemann \& Khalil \citep{HeidemannKhalil} and the critical points obtained between the two techniques are absolutely similar. In view of this fact, and for the sake of conciseness, numerical comparisons will not be presented here.}

\section{Conclusions} \label{sec:conclusions}

In this work we analyzed the calculation of critical points in binary mixtures using the formulation of \citet{HeidemannKhalil} by means 
of  a numerical algorithm based on the methodology proposed by \citet{bib:malta1996numerical}. The numerical strategy was applied in two binary mixtures, allowing the numerical calculation of the critical points in a robust and accurate way.

The results obtained using the numerical inversion of functions produced approximations that are virtually the same to those obtained by the classical algorithm---with a double-loop structure---of \citet{HeidemannKhalil}. A comparison with the results produced by a damped Newton algorithm, proposed by \citet{Dimitrakopoulos}, and with the algorithm proposed by \citet{nichita2010} also indicates very accurate critical coordinates.

Furthermore, the steps implemented in the methodology (the construction of critical curves, the generation of the bank of solved points and, finally, the inversion process itself) were also useful to throw light on the behavior of the functions that characterize the critical conditions.

\section*{Acknowledgements}
The authors acknowledge the financial support of Brazilian Agencies CAPES, CNPq and FAPERJ.


\bibliographystyle{apacite}
\bibliography{bibliography.bib}

\section*{Appendix}
\textcolor{black}{According to \cite{HeidemannKhalil}, a phase under conditions defined by the state $(T_0, V_0, n_{1,0}, n_{2,0}, \dots, n_{c,0})$---where $T_0$ and $V_0$ are, respectively, the system temperature and volume, and $n_{i,0}$, with $i=1,2,\dots, c$, are the number of mols of the components of the mixture---is stable if, for any isothermal variation, the following constraint is satisfied:}
\textcolor{black}{
\begin{equation} \label{funcao_D}
 \left[ A - A_0 + P_0 \Delta V- \sum\limits_{i = 1}^{c} \mu_{i,0}\Delta  n_i\right]_{T_0} >0\;, 
\end{equation}}
\textcolor{black}{being $(T_0, V, n_{1}, n_{2}, \dots, n_{c})$ the new state, $\Delta V=V-V_0$ the volume variation, $\Delta n_i=n_i-n_{i,0}$  the difference between the number of mols in the new state and the initial state, and $\mu_{i,0}$ representingthe chemical potential of the $i$-th component at the initial state. The Helmholtz free energy at the new state is represented by $A$, while $A_0$ and $P_0$ denote, respectively, the Helmholtz free energy and the system pressure at the initial state.}

\textcolor{black}{It must be emphasized that, given $k \neq 0$, variations of the form $\Delta V = kV_0 $ and $ \Delta  n_i = kn_{i,0} $,
 for $ i=1,\dots, c $, cannot be considered a change in phase, since the mole fractions and density will be constant. Therefore, the pressure and the chemical potentials do not undergo variations either. In order to avoid this, \cite{HeidemannKhalil} adopt the condition $ \Delta V = 0 $ and, then, Equation~(\ref{funcao_D}) can be simplified to}
\textcolor{black}{
\begin{equation*}
\left[ A - A_0 - \sum\limits_{i = 1}^{c} \mu_{i,0}\Delta n_i\right]_{T_0, V_0} >0\;.
\end{equation*}}
\textcolor{black}{
\noindent Once $\mu_i=\left(\dfrac{\partial A}{\partial n_i}\right)_{T,V,n_{l\neq i}}$, then, 
\begin{equation}
\left[ A - A_0 - \sum\limits_{i = 1}^{c} \left(\dfrac{\partial A}{\partial n_i}\right)_{T_0,V_0,n_{0,l\neq i}}\Delta n_i\right]_{T_0, V_0} >0\;.
\label{dd}
\end{equation}}

\textcolor{black}{
Consider a Taylor series expansion of Helmholtz free energy $A$ around the initial state (state $0$), under constant volume ($V$):
\begin{equation}\label{eq1_criticalidade}
\begin{split}
A = A_0 + \sum\limits_{i=1}^{c}&\left(\dfrac{\partial A}{\partial n_i}\right)_{T_0,V_0,n_{0,l\neq i}}\Delta n_i+\dfrac{1}{2}\sum\limits_{j=1}^{c}\sum\limits_{i=1}^{c}\left(\dfrac{\partial^2 A}{\partial n_i \partial n_j}\right)_{T_0,V_0,n_{0,l\neq j,i}}\Delta n_i \Delta n_j + \\
\dfrac{1}{6}&\sum\limits_{k=1}^{c}\sum\limits_{j=1}^{c}\sum\limits_{i=1}^{c}\left(\dfrac{\partial^ 3 A}{\partial n_i\partial n_j\partial n_k}\right)_{T_0,V_0,n_{0,l\neq k,j,i}}\Delta n_i\Delta n_j \Delta n_k + O(\Delta n^4)\;,
\end{split}
\end{equation}}
\textcolor{black}{
\noindent where $O(\Delta n^4)$ is the residual term of the Taylor series after the third-order term, and the subscript $n_{0,l\neq k,j,i}$ refers to the number of mols for all components, except $i$, $j$ or $k$.}

\textcolor{black}{
This expansion together with Equation~(\ref{dd})  implies, without loss of generality, in:
\begin{equation}\label{exp_taylor}
\begin{split}
\dfrac{1}{2}\sum\limits_{j=1}^{c}\sum\limits_{i=1}^{c}&\left(\dfrac{\partial^2 A}{\partial n_i \partial n_j}\right)_{T_0,V_0,n_{0,l\neq j,i}}\Delta n_i \Delta n_j + \\
\dfrac{1}{6}&\sum\limits_{k=1}^{c}\sum\limits_{j=1}^{c}\sum\limits_{i=1}^{c}\left(\dfrac{\partial^ 3 A}{\partial n_i\partial n_j\partial n_k}\right)_{T_0,V_0,n_{0,l\neq k,j,i}}\Delta n_i\Delta n_j \Delta n_k + O(\Delta n^4)>0\;.
\end{split}
\end{equation}}

\textcolor{black}{Since a critical point is determined by the stability condition limit, the quadratic and cubic terms must vanish and, then, the stability is guaranteed by higher-order terms.}

\end{document}